\documentclass[10pt,english,floatfix,nofootinbib,superscriptaddress,aps,prd,preprint,twocolumn,showkeys]{revtex4}
\usepackage[utf8]{inputenc}
\usepackage{float}
\usepackage{array}
\usepackage{lipsum}
\usepackage{bbm}
\usepackage{dsfont}
\usepackage{graphicx}
\usepackage{amsmath}
\usepackage{tikz}
\usepackage{multirow}
\usepackage{braket}
\usepackage{bbm}
\usetikzlibrary{quotes,angles}
\usetikzlibrary{arrows} 
\usetikzlibrary{decorations.markings}
\usepackage{graphicx}
\usepackage[english]{babel}
\usepackage{color}
\usepackage{subfig}
\usepackage{caption}
\usepackage{tensor}
\usepackage{booktabs}
\usepackage{esint}
\usepackage[dvips]{epsfig}
\usepackage[dvips]{graphicx}
\usepackage{float}
\usepackage{units}
\usepackage{textcomp}
\usepackage{mathrsfs}
\usepackage{amsmath}
\usepackage{cancel}
\usepackage{amssymb}
\usepackage{amsbsy}
\usepackage{amsfonts}
\usepackage{amssymb,mathrsfs,xcolor}
\usepackage{esint}
\usepackage{array}
\usepackage{graphicx}
\usepackage{hyperref}
\usepackage{enumerate}
\hypersetup{
    colorlinks,
    citecolor=blue,
    filecolor=green,
    linkcolor=purple,
    urlcolor=red,
}
\usepackage{slashed}
\usepackage{hyperref}
\hypersetup{colorlinks,breaklinks,
		  citecolor=[rgb]{0,0.0,1.0},
            urlcolor=[rgb]{0.0,0.0,1.0},
            linkcolor=[rgb]{0,0.5,0.9}}

\usepackage{changes}

\begin{document}

\title{The neutrino flavor oscillations in the static and spherically symmetric black-hole-like wormholes}

\author{Yuxuan Shi}
\email{shiyx2280771974@Mail.com}
\affiliation{Department of Physics, East China University of Science and Technology, Shanghai 200237, China}

\author{Hongbo Cheng}
\email{hbcheng@ecust.edu.cn}
\affiliation{Department of Physics, East China University of Science and Technology, Shanghai 200237, China}
\affiliation{The Shanghai Key Laboratory of Astrophysics, Shanghai, 200234, China}

\begin{abstract}
We study the effects of neutrino lensing induced by a Damour-Solodukhin wormhole on the neutrino oscillation. We derive and calculate the flavour transition probabilities in the presence of Damour-Solodukhin factor $\Lambda$ as a shift in the massive source to show that the neutrino flavour oscillation is also sensitive not only to the sign of difference between the squared masses but also to the individual mass of neutrinos in both the two-flavour and the three-flavour cases. As a consequence of parameter $\Lambda$ within a region, a series of curves of probability function versus the azimuthal angle $\phi$ with definite masses of neutrino can be plotted and their shapes resemble each other in the case of two-flavoured neutrinos and of three-flavoured ones. In view of the probability functions due to the wormhole, we reveal that the contribution of the factor $\Lambda$ is novel. Based on our analytical and numerical discussions on the probability expressions, the difference of the neutrino flavour oscillation arising from the shift in the wormhole factor $\Lambda$ is detectable.  It is crucial that the $\Lambda$ as deviation from the black holes can change the shapes of the curves greatly, in the case of three-flavoured neutrinos in particular. The detailed comparisons can be made among our estimations depicted in the figures for neutrino oscillations and the measurements from the detector, which open a new window for judging whether the remote star as lens is black-hole-like wormhole or spherically symmetric black hole and further the wormhole factor $\Lambda$ can be estimated.
\end{abstract}

\maketitle


\section{Introduction}

A lot of significant measurements and theoretical predictions have been reported during the research on compact stars. The black holes, or wormholes, as a kind of strange celestial bodies are theoretical solutions to the Einstein equations \cite{penrose1965gravitational, chandrasekhar1998mathematical,abbott2016observation,abbott2019gwtc,visser1995lorentzian,morris1988wormholes}. One key experiment like the Event Horizon Telescope (EHT) makes it possible to deepen the exploration of the black hole core \cite{collaboration2019first}. At the throat of any static and spherically symmetric travable wormhole, the weak energy condition cannot be kept under the general relativity, but the wormhole solutions to the field equation under the weak energy condition can be found in the alternative gravity \cite{morris1988wormholes, shaikh2016wormholes}. Based on the theoretical analysis and astronomical observation, the black holes and wormholes have their own properties: an event horizon for black holes and a throat for wormholes \cite{chandrasekhar1998mathematical,abbott2016observation,abbott2019gwtc, visser1995lorentzian,morris1988wormholes}. As mentioned above, more achievements have been made on the black hole, but these special celestial bodies need to be deeply investigated \cite{penrose1965gravitational, chandrasekhar1998mathematical,abbott2016observation, abbott2019gwtc,visser1995lorentzian,morris1988wormholes,collaboration2019first}. The research on the wormholes can help to further the investigation. It is necessary to probe the properties of wormholes in various directions. It is worth listing the works that contributed to the wormholes that the electromagnetic rays pass through in the strong gravitational field, like deflection angles \cite{chetouani1984geometrical}, visualizations \cite{muller2004visual, james2015visualizing}, gravitational lensing \cite{perlick2004exact,nandi2006gravitational, tsukamoto2017light,shaikh2019novel}, wave optics \cite{shaikh2019erratum}, etc.. The authors of Ref.\cite{shaikh2019strong} focused on the orbits of stars or pulsars around the black holes at the centre of the galaxy to describe the relevant wormholes. As a starting point, we select the wormholes that can recover to be the black holes by neglecting the shift \cite{visser1995lorentzian,nambu2019wave}. This kind of wormhole, named black hole foils, has no event horizon but a throat \cite{nambu2019wave}. Some physicists performed the research on the so-called black-hole-like wormholes to distinguish the globally static wormholes from the black holes that resemble them \cite{nambu2019wave}. It was shown that Hawking's radiation from the wormholes is too weak to be measured \cite{visser1995lorentzian, nambu2019wave}. The author of Ref.\cite{dai2019observing} pointed out that a tidal force acting on a falling body near the Damour-Solodukhin wormhole is greater than that near the Schwarzschild black hole by comparison. Some efforts have been contributed to the accretion disks and images of the Ker-typed wormholes \cite{simonetti2021sensitive, bambi2021astrophysical}. The wormholes were thought to be black hole foils \cite{damour2007wormholes}, and their quasinormal modes were derived and calculated \cite{damour2007wormholes, lemos2008black}. The accretion disk, the signature from potentials and throats, and the shadow image of the Damour-Solodukhin wormhole were investigated \cite{karimov2019accretion, paul2020observational, volkel2018wormhole, ovgun2021quasinormal, ovgun2018light, ovgun2018reply, ovgun2019deflection, nandi2018strong, amir2019shadow}. Some physicists paid attention to the gravitational-wave echoes by the wormholes \cite{bueno2018echoes, cardoso2019gravitational, li2019mixing, bronnikov2020echoes, tsukamoto2020high}. The centre-of-mass energy for the head-on collision of two test particles around the Damour-Solodukhin wormhole, whose metric has a tiny deviation from the corresponding black holes \cite{tsukamoto2020high}. We studied the energy deposition rate from the neutrino pair annihilation around the Damour-Solodukhin wormhole, thought to be a mimicker of the Schwarzschild black hole, to indicate that the wormhole'saccretion disk can become a source of gamma-ray burst \cite{shi2023gamma}.

The fundamental characteristics of neutrinos have attracted a lot of attention for a long time \cite{capozzi2014status, de2018status, esteban2019global, particle2020review, esteban2020fate, an2012observation}. The neutrinos properties involve at least three flavors of weakly interacting neutrinos, while the mass eigenstates of neutrinos are not the same as their flavor eigenstates. The analysis of the neutrino oscillation phenomena opens a window to explore these characteristics \cite{capozzi2014status, de2018status, esteban2019global}. According to the theoretical research and corresponding measurements on the neutrino oscillations in flat spacetime, it was found that the neutrino oscillation probabilities depend on the difference of the square of neutrino masses like $|\Delta m_{21}^2|$, $|\Delta m_{31}^2|$ and $|\Delta m_{23}^2|$, where $\Delta m_{ij}^2=m_i^2-m_j^2$, but not the individual neutrino masses \cite{tanabashi2018review}. In view of the previous works, the phenomenon of neutrino flavor oscillation in curved spacetime can provide information about the absolute neutrino masses and the oscillation probabilities, which can be utilized to show the background structure of the gravitational objects in contact with the astrophysical measurements. The discussions on neutrino propagation were generalized to the background of gravitational sources \cite{ahluwalia1996gravitationally, ahluwalia1996interpretation, grossman1997flavor, bhattacharya1999gravitationally, luongo2011neutrino, geralico2012neutrino, koutsoumbas2020neutrino, swami2020signature}. It is useful to investigate the gravitational potential of neutrino propagation along the geodesics \cite{cardall1997neutrino, fornengo1997gravitational}. The gravity can also lead to the lensing of neutrino oscillation probabilities, meaning that the neutrino with different trajectories converges at a common point in the vicinity of the mass source \cite{swami2020signature}. There have been more efforts contributed to the neutrino flavor oscillation probability at the focus \cite{swami2020signature, crocker2004neutrino, alexandre2018black, dvornikov2020spin}. Swami furthered the analysis on the neutrino phase in a rotating spacetime to show that the corrections from the angular momentum making the rotation to the oscillation probabilities are significant and the corrected probabilities are considerable when the celestial body is more sun-sized \cite{swami2022neutrino}. The effects of gravitational lensing on the two or three flavors of neutrino oscillations were derived and calculated in the spacetime specified by the deformation parameter $\gamma$, leading the background of the gravitational body to be static, axially-symmetric and asymptotically flat \cite{chakrabarty2022effects}. It was shown that the factor $\gamma$ can amend the neutrino oscillation and the oscillation probabilities have something to do with the individual masses of neutrinos \cite{chakrabarty2022effects}.

It is significant to research the neutrino flavor oscillation in the static and spherically symmetric black-hole-like wormhole. In several kinds of curved spacetime, neutrino oscillations were explored. It is important that the spacetime structure give rise to an influence on the oscillation probabilities while the probabilities depend on the neutrino masses. As a kind of special compact gravitating object, the wormholes need to be studied in various directions. Fewer researchers have addressed the process of neutrino conversion around the wormholes. We considered the neutrino pair annihilation around the Damour-Solodukhin wormhole involving the shift from the similar black holes to show that the deviation may reduce the emitted power slightly, but these static and spherically symmetric black-hole-like wormholes in this case can become sources of gamma-ray burst \cite{shi2023gamma}. There remains a need for a new and potential method that can distinguish the wormholes from black holes. Here we are going to investigate the phase of neutrino oscillation around the black-hole-like wormholes to exhibit the effects of the positive dimensionless parameter $\Lambda$ causing the metric to be different from the Schwarzschild's ones on the neutrino oscillation probabilities. We can compare the theoretical evaluation of the neutrino oscillation with the measurements by the detector on the earth to wonder how the factor $\Lambda$ specifies the massive source metric. We are also going to find the dependence of the oscillation probabilities in the wormhole spacetime on the absolute neutrino masses and the sign of the mass-squared difference. This paper is organized as follow. First, we consider the neutrino oscillations in the static and spherically symmetric black-hole-like wormholes to find the oscillation probabilities as functions of the individual neutrino masses and the neutrino mass-squared difference under the influence of deviation from the standard black hole. It is necessary to discuss the probabilities and compare the results with the detection in the background of a gravitational source. The process can be thought of as an effective method to discern the black hole and the black-hole-like wormhole. The discussion and conclusions are presented in the end.

\section{Phase and probability of neutrino oscillation}

At starting step, we revisit the geodesic for a particle motion in the Damour-Solodukhin spacetime \cite{nambu2019wave, simonetti2021sensitive}. The metric of static and spherically symmetric spacetime can be written as \cite{nambu2019wave},
\begin{align}
\label{ds2}
\mathrm{d}s^2
&= g_{\mu\nu}\mathrm{d}x^{\mu}\mathrm{d}x^{\nu}\notag\\
&= \left[f(r)+\Lambda^2\right]\mathrm{d}t^2-\dfrac{\mathrm{d}r^2}{f(r)}-r^2\left(\mathrm{d}\theta^2+\sin^2\theta\mathrm{d}\varphi^2\right),
\end{align}
where the part of a component is
\begin{align}
\label{fr}
f(r) = 1-\dfrac{2M}{r}
\end{align}
with the Newton constant $G$ and the mass $M$. The positive dimensionless parameter $\Lambda$ indicates deviations from the Schwarzschild metric \cite{nambu2019wave}. $\Lambda\ll1$ is the constraint. It should be pointed out that this kind of metric (\ref{ds2}) has the throat at $r=2M$ that joins two isometric, asymeptotically flat regions and can describe the wormhole \cite{nambu2019wave}. In the spherically symmetric spacetime with the description of metric (\ref{ds2}), the Lagrangian for the motion of the neutrinos in the $k$-th eigenstate is written as \cite{nambu2019wave}
\begin{align}
\label{lagrangian}
\mathcal{L}
&= \dfrac{1}{2}m_{k}g_{\mu\nu}\dfrac{\mathrm{d}x^{\mu}}{\mathrm{d}\tau}\dfrac{\mathrm{d}x^{\nu}}{\mathrm{d}\tau}\notag\\
&= \dfrac{1}{2}m_{k}\left[f(r)+\Lambda^2\right]\left(\dfrac{\mathrm{d}t}{\mathrm{d}\tau}\right)^2-\dfrac{1}{2}\dfrac{m_{k}}{f(r)}\left(\dfrac{\mathrm{d}r}{\mathrm{d}\tau}\right)^2\notag\\
&\quad-\dfrac{1}{2}m_{k}r^2\left(\dfrac{\mathrm{d}\theta}{\mathrm{d}\tau}\right)^2-\dfrac{1}{2}m_{k}r^2\sin^2\theta\left(\dfrac{\mathrm{d}\varphi}{\mathrm{d}\tau}\right)^2
\end{align}
with proper time $\tau$ and $m_{k}$, the mass of the $k$-th eigenstate. The canonical conjugate momentum to the coordinate $x^{\mu}$ of the particle is $p_{\mu}=\frac{\partial\mathcal{L}}{\partial\frac{\mathrm{d}x}{\mathrm{d}\tau}}$ and the nonzero  components of the 4-momentum for the particles moving on the equatorial plane $\theta=\frac{\pi}{2}$ in the wormhole spacetime are \cite{nambu2019wave, chakrabarty2022effects}
\begin{align}
\label{momentum}
p^{(k)t}&=m_{k}\left[f(r)+\Lambda^2\right]\dfrac{\mathrm{d}t}{\mathrm{d}\tau}=E_{k},\notag\\
p^{(k)r}&=\dfrac{m_{k}}{f(r)}\dfrac{\mathrm{d}r}{\mathrm{d}\tau},\\
p^{(k)\varphi}&=m_{k}r^2\dfrac{\mathrm{d}\varphi}{\mathrm{d}\tau}=J_{k}.\notag
\end{align}
The mass of the $k$-th eigenstate satisfies the mass-shell relation \cite{nambu2019wave, cardall1997neutrino, fornengo1997gravitational},
\begin{align}
\label{mass-shell}
m_{k}^2 = g_{\mu\nu}p^{(k)\mu}p^{(k)\nu}.
\end{align}

\subsection{Radial case}
It is significant to provide some effective ways to characterize the neutrino flavour oscillation. First we gain an insight into the radial propagation of neutrinos on the equatorial plane in the Damour-Solodukhin spacetime. We start by looking into the situation of neutrinos propagating radially, $\mathrm{d}\varphi=0$. A neutrino moving radially along a light-ray trajectory changes its phase to \cite{chakrabarty2022effects,godunov2011neutrino}
\begin{align}
\label{phi_r}
\Phi_k=\int_{\left(t_{S},\bm{x}_{S}\right)}^{\left(t_{D},\bm{x}_{D}\right)}\left[E_k\left(\dfrac{\mathrm{d}t}{\mathrm{d}r}\right)_0+p_k(r)\right]\mathrm{d}r,
\end{align}
where $\left(t_{S},\bm{x}_{S}\right)$ and $\left(t_{D},\bm{x}_{D}\right)$ stand for the neutrino's and detector's respective sources. We get the light-ray differential as \cite{chakrabarty2022effects}
\begin{align}
\label{diff}
\left(\dfrac{\mathrm{d}t}{\mathrm{d}r}\right)_0=\dfrac{E_0}{p_0(r)}\dfrac{g_{rr}}{g_{tt}},
\end{align}
where $E_0$ is the energy of a massless particle. Using the mass-shell relation Eq.(\ref{mass-shell}), we can obtain
\begin{align}
\label{p0_r}
p_0(r)=\pm E_0\sqrt{\dfrac{g_{rr}}{g_{tt}}}
\end{align}
and
\begin{align}
\label{pk_r}
p_k(r)=\pm\sqrt{E_k^2\dfrac{g_{rr}}{g_{tt}}-g_{rr}m_k^2}.
\end{align}
In view of the momenta specified by Eqs.(\ref{momentum}) and (\ref{diff}-\ref{pk_r}), we aim to the phase for radial neutrino emission \cite{cardall1997neutrino,fornengo1997gravitational,chakrabarty2022effects,godunov2011neutrino},
\begin{align}
\Phi_k=\pm\int_{\left(t_{S},\bm{x}_{S}\right)}^{\left(t_{D},\bm{x}_{D}\right)}E_k\sqrt{\dfrac{g_{rr}}{g_{tt}}}\left[-1+\sqrt{1-g_{tt}\dfrac{m_k^2}{E_k^2}}\right]\mathrm{d}r.
\end{align}
The square-root under the bracket could be expanded to the order $\mathcal{O}\left(\frac{m_k^2}{E_k^4}\right)$, following \cite{godunov2011neutrino},
\begin{align}
\label{phi1}
\Phi_k&=\pm\biggl(\dfrac{m_k^2}{2E_k}\int_{\left(t_{S},\bm{x}_{S}\right)}^{\left(t_{D},\bm{x}_{D}\right)}\sqrt{g_{tt}g_{rr}}\mathrm{d}r\notag\\
&\quad+\dfrac{m_k^4}{8E_k^3}\int_{\left(t_{S},\bm{x}_{S}\right)}^{\left(t_{D},\bm{x}_{D}\right)}g_{tt}^{\frac{3}{2}}\sqrt{g_{rr}}\mathrm{d}r\biggr).
\end{align}
The function in the Eq.(\ref{phi1}) can be expanded to order $\mathcal{O}\left(\frac{M^2}{r^2}\right)$. The first term becomes into
\begin{align}
\label{int1}
&\quad\int_{\left(t_{S},\bm{x}_{S}\right)}^{\left(t_{D},\bm{x}_{D}\right)}\sqrt{g_{tt}g_{rr}}\mathrm{d}r\notag\\
&\simeq\int_{\left(t_{S},\bm{x}_{S}\right)}^{\left(t_{D},\bm{x}_{D}\right)}\sqrt{\dfrac{f(r)+\Lambda^2}{f(r)}}\mathrm{d}r\notag\\
&=\sqrt{1+\Lambda^2}\left(r_D-r_S\right)+\dfrac{\Lambda^2M}{\sqrt{1+\Lambda^2}}\ln\left(\dfrac{r_D}{r_S}\right)\notag\\
&\quad+\dfrac{\Lambda^2\left(4+3\Lambda^2\right)}{2\left(1+\Lambda^2\right)^{\frac{3}{2}}}\dfrac{M^2\left(r_D-r_S\right)}{r_D r_S}.
\end{align}
Similarly, the second term in Eq.(\ref{phi1}) becomes
\begin{align}
\label{int2}
&\quad\int_{\left(t_{S},\bm{x}_{S}\right)}^{\left(t_{D},\bm{x}_{D}\right)}g_{tt}^{\frac{3}{2}}\sqrt{g_{rr}}\mathrm{d}r\notag\\
&\simeq\int_{\left(t_{S},\bm{x}_{S}\right)}^{\left(t_{D},\bm{x}_{D}\right)}\dfrac{\left[f(r)+\Lambda^2\right]^{\frac{3}{2}}}{\sqrt{f(r)}}\mathrm{d}r\notag\\
&=\left(1+\Lambda^2\right)^{\frac{3}{2}}\left(r_D-r_S\right)+\sqrt{1+\Lambda^2}\left(\Lambda^2-2\right)M\ln\left(\dfrac{r_D}{r_S}\right)\notag\\
&\quad+\dfrac{3\Lambda^4}{2\sqrt{1+\Lambda^2}}\dfrac{M^2\left(r_D-r_S\right)}{r_D r_S}.
\end{align}
The phase expression for the radial case expanded to the order $\mathcal{O}\left(\frac{m_k^2}{E_k^4}\right)$ is obtained by combining Eqs.(\ref{int1}) and (\ref{int2}) into Eq.(\ref{phi1}),
\begin{align}
\label{phi2}
\Phi_k&=\pm\Biggl\{\dfrac{m_k^2}{2E_k}\Biggl[\sqrt{1+\Lambda^2}\left(r_D-r_S\right)+\dfrac{\Lambda^2M}{\sqrt{1+\Lambda^2}}\ln\left(\dfrac{r_D}{r_S}\right)\notag\\
&\quad+\dfrac{\Lambda^2\left(4+3\Lambda^2\right)}{2\left(1+\Lambda^2\right)^{\frac{3}{2}}}\dfrac{M^2\left(r_D-r_S\right)}{r_D r_S}\Biggr]\notag\\
&\quad+\dfrac{m_k^4}{8E_k^3}\Biggl[\left(1+\Lambda^2\right)^{\frac{3}{2}}\left(r_D-r_S\right)\notag\\
&\quad+\sqrt{1+\Lambda^2}\left(\Lambda^2-2\right)M\ln\left(\dfrac{r_D}{r_S}\right)\notag\\
&\quad+\dfrac{3\Lambda^4}{2\sqrt{1+\Lambda^2}}\dfrac{M^2\left(r_D-r_S\right)}{r_D r_S}\Biggr]\Biggr\}.
\end{align}
Let $\Lambda=0$ and considering $\Delta\Phi_{ij}=\Phi_i-\Phi_j$, it will revert to the Schwarzschild scenario,
\begin{align}
\Delta\Phi_{ij}|_{\Lambda=0}&=\pm\Biggl[\dfrac{\Delta m_{ij}^2}{2E_k}\left(r_D-r_S\right)+\dfrac{\Delta m_{ij}^4}{8E_k^3}\left(r_D-r_S\right)\notag\\
&-\dfrac{\Delta m_{ij}^4}{8E_k^3}\ln\left(\dfrac{r_D}{r_S}\right)\Biggr],
\end{align}
where $\Delta m_{ij}^2=m_i^2-m_j^2$ and $\Delta m_{ij}^4=m_i^4-m_j^4$. Our findings aligned with the result in Ref.\cite{godunov2011neutrino}. It is obvious that the terms consisting of expression of phase or phase shift contain the Damour-Solodukhin factor $\Lambda$ according to Eq.(\ref{phi1}) and Eq.(\ref{phi2}). This factor changes the phase drastically because the vanishing factor leads the two terms belonging to the expressions to disappear. It is noted that most of neutrinos just move along one radial direction in the universe.

\subsection{Non-radial case}
In curved spacetime, the neutrino flavor oscillation has been studied for the plane wave approximation within the frame of weak gravity \cite{swami2020signature,cardall1997neutrino}. In weak interactions, neutrinos are always denoted and detected in flavor eigenstates like \cite{pontecorvo1957inverse, maki1962remarks, pontecorvo1967neutrino}, 
\begin{align}
\label{eigenstate}
\ket{\nu_{\alpha}}=\sum_{i=1}^{3}U_{\alpha i}^{*}\ket{\nu_{i}},
\end{align}
where $\alpha=e,\mu,\nu$, the three flavors of neutrinos. The mass eigenstate are denoted as $\ket{\nu_{i}}$. $U$ identified with the leptonic mixing matrix is $3\times3$ unitary matrix and relates the flavor eigenstate to the mass eigenstate \cite{esteban2019global}. The wave functions can be used to describe the neutrino mass eigenstate and the neutrino propagation from one spacetime point to another one. It is adequate to choose the coordinates $\left(t_{S},\bm{x}_{S}\right)$ and $\left(t_{D},\bm{x}_{D}\right)$ for a source $S$ and a detector $D$ respectively \cite{nambu2019wave, fornengo1997gravitational, pontecorvo1957inverse, maki1962remarks, pontecorvo1967neutrino, stodolsky1979matter}. The wave function can be written as \cite{nambu2019wave, fornengo1997gravitational, pontecorvo1957inverse, maki1962remarks, pontecorvo1967neutrino, stodolsky1979matter},
\begin{align}
\label{DtoS}
\ket{\nu_{i}\left(t_{D},\bm{x}_{D}\right)}=\exp\left(-\mathrm{i}\Phi_{i}\right)\ket{\nu_{i}\left(t_{S},\bm{x}_{S}\right)},
\end{align}
where the phase is 
\begin{align}
\label{phase}
\Phi_{k}=\int_{\left(t_{S},\bm{x}_{S}\right)}^{\left(t_{D},\bm{x}_{D}\right)}g_{\mu\nu}p^{(i)\mu}\mathrm{d}x^{\nu}.
\end{align}

Now we return to the flavor oscillation in the neutrino propagation between the source where the neutrino is generated and the detector. The oscillation probability for $\nu_{\alpha}\to\nu_{\beta}$ at the detection point is obtained \cite{nambu2019wave, fornengo1997gravitational, pontecorvo1957inverse, maki1962remarks, pontecorvo1967neutrino},
\begin{align}
\label{Pab}
P_{\alpha\beta}
&= |\left\langle \nu_{\beta}|\nu_{\alpha}\left(t_D, \bm{x}_D\right)\right\rangle|^2\notag\\
&=\sum_{i,j=1}^3U_{\beta i}U_{\beta j}^*U_{\alpha j}U_{\alpha i}^*\exp[-\mathrm{i}(\Phi_{i}-\Phi_{j})].
\end{align}

We discuss the motion of neutrinos limited on the $\theta=\frac{\pi}{2}$-plane subject to the gravitational field of the Damour-Solodukhin wormhole. We substitute the wormhole metric (\ref{fr}) into the phase (\ref{phase}) with the help of momentum components (\ref{momentum}) to get \cite{cardall1997neutrino, fornengo1997gravitational},
\begin{widetext}
\begin{align}
\label{Phi}
\Phi_{k}
&= \int_{\left(t_{S},\bm{x}_{S}\right)}^{\left(t_D, \bm{x}_D\right)}g_{\mu\nu}p^{(k)\mu}\mathrm{d}x^{\nu}\notag\\
&= \int_{\left(t_{S},\bm{x}_{S}\right)}^{\left(t_D, \bm{x}_D\right)}\left[E_k\mathrm{d}t-p^{(k)r}\mathrm{d}r-J_k\mathrm{d}\varphi\right]\mathrm{d}r\notag\\
&\simeq \pm\Biggl[\dfrac{m_k^2}{2E_0}\int_{r_S}^{r_D}\dfrac{\sqrt{f(r)+\Lambda^2}}{\sqrt{f(r)}}\left\{1-\dfrac{b^2}{r^2}\left[f(r)+\Lambda^2\right]\right\}^{-\frac{1}{2}}\mathrm{d}r\notag\\
&\quad+\dfrac{m_k^4}{8E_0^3}\int_{r_S}^{r_D}\dfrac{1}{\sqrt{\left[f(r)+\Lambda^2\right]f(r)}}\left\{1-\dfrac{b^2}{r^2}\left[f(r)+\Lambda^2\right]\right\}^{-\frac{1}{2}}\mathrm{d}r\Biggr]\notag\\
&= \pm\dfrac{m_k^2}{2E_0}\Biggl\{\sqrt{1+\Lambda^2}\left[\sqrt{r_D^2-b^2\left(1+\Lambda^2\right)}-\sqrt{r_S^2-b^2\left(1+\Lambda^2\right)}\right]\notag\\
&\quad+\dfrac{M}{\sqrt{1+\Lambda^2}}\left[\dfrac{r_D}{\sqrt{r_D^2-b^2\left(1+\Lambda^2\right)}}-\dfrac{r_S}{\sqrt{r_S^2-b^2\left(1+\Lambda^2\right)}}\right]-\dfrac{M\Lambda^2}{\sqrt{1+\Lambda^2}}\ln\left[\dfrac{r_D-\sqrt{r_D^2-b^2\left(1+\Lambda^2\right)}}{r_S-\sqrt{r_S^2-b^2\left(1+\Lambda^2\right)}}\right]\Biggr\}\notag\\
&\quad\pm\dfrac{m_k^4}{8E_0^3}\Biggl\{\sqrt{\dfrac{r_D^2}{1+\Lambda^2}-b^2}-\sqrt{\dfrac{r_S^2}{1+\Lambda^2}-b^2}+\dfrac{M}{\left(1+\Lambda^2\right)^{\frac{3}{2}}}\left[\dfrac{r_D}{\sqrt{r_D^2-b^2\left(1+\Lambda^2\right)}}-\dfrac{r_S}{\sqrt{r_S^2-b^2\left(1+\Lambda^2\right)}}\right]\notag\\
&\quad-\dfrac{M\left(2+\Lambda^2\right)}{\left(1+\Lambda^2\right)^{\frac{3}{2}}}\ln\left[\dfrac{r_D-\sqrt{r_D^2-b^2\left(1+\Lambda^2\right)}}{r_S-\sqrt{r_S^2-b^2\left(1+\Lambda^2\right)}}\right]\Biggr\},
\end{align}
\end{widetext}
where $E_0=\sqrt{E_k^2-m_k^2}$, the average energy of the relativistic neutrinos originated at the source, and $b$ is the impact factor \cite{nambu2019wave}. It is pointed that the phase of the k-th mass eigenstate of neutrinos with the description of Eq.(\ref{Phi}) will recover to be that in the case of Schwarzschild black hole when $\Lambda=0$ \cite{cardall1997neutrino, fornengo1997gravitational}. According to the momenta of the neutrinos propagating on the equatorial plane in the black-hole-like wormhole spacetime in terms of Eq.(\ref{momentum}), the angular component of momentum can be found that $J_{k}=\frac{E_{k}}{f(r)+\Lambda^{2}r^{2}\frac{\mathrm{d}\varphi}{\mathrm{d}t}}$ \cite{nambu2019wave,chakrabarty2022effects}. Further the wormhole-corrected momentum component become $\lim_{r\to\infty}J_{k}=\frac{E_{k}}{1+\Lambda^{2}}bv_{k}^{\infty}$ because of the neutrinos remote from the Damour-Solodukhin wormhole, where $\lim_{r\to\infty}r^{2}\frac{\mathrm{d}\varphi}{\mathrm{d}t}=bv^{\infty}$ with the impact factor and asymptotic velocity \cite{nambu2019wave,cardall1997neutrino, fornengo1997gravitational}. Based on the wormhole metric (\ref{ds2}), the Damour-Solodukhin parameter just shifts one component of the corresponding black hole metric a little, but the spacetime structures including the asymptotic behavour of metric have been changed \cite{damour2007wormholes}. The phase (\ref{phi1}) declares that all of the terms relate to the wormhole shift. When $\Lambda=0$, the last two parts will be equal to zero. The corrections from the wormholes to the phase are considerably evident. During the propagation, there exists the closest point of approach at $r=r_{0}$ for the propagating neutrinos. Under the weak field approximation, the shortest distance $r_{0}$ as a solution to the orbital equation of neutrino is obtained \cite{chandrasekhar1998mathematical},
\begin{align}
\label{r0}
r_0 \simeq b\sqrt{1+\Lambda^2}-\dfrac{M}{1+\Lambda^2}.
\end{align}
When the neutrinos move from the source, pass through the point of the closest approach and get to the detector, their phase can be expressed as according to the Eq.(\ref{Phi}) and Eq.(\ref{r0}) \cite{cardall1997neutrino, fornengo1997gravitational},
\begin{widetext}
\begin{align}
\label{PhiS0D}
&\quad\Phi_k\left(r_S\to r_0\to r_D\right)\notag\\
&\simeq \dfrac{m_k^2}{2E_0}\Biggl\{\dfrac{Mb}{\sqrt{r_D^2-b^2\left(1+\Lambda^2\right)}}+\dfrac{Mb}{\sqrt{r_S^2-b^2\left(1+\Lambda^2\right)}}\notag\\
&\quad+\sqrt{1+\Lambda^2}\left[\sqrt{r_D^2-b^2\left(1+\Lambda^2\right)}+\sqrt{r_S^2-b^2\left(1+\Lambda^2\right)}\right]\notag\\
&\quad+\dfrac{M}{\sqrt{1+\Lambda^2}}\left(\dfrac{r_D-b\sqrt{1+\Lambda^2}}{r_D+b\sqrt{1+\Lambda^2}}+\sqrt{\dfrac{r_S-b\sqrt{1+\Lambda^2}}{r_S+b\sqrt{1+\Lambda^2}}}\right)\notag\\
&\quad-\dfrac{M\Lambda^2}{\sqrt{1+\Lambda^2}}\ln\left[\dfrac{\left(r_D-\sqrt{r_D^2-b^2\left(1+\Lambda^2\right)}\right)\left(r_S-\sqrt{r_S^2-b^2\left(1+\Lambda^2\right)}\right)}{b^2\left(1+\Lambda^2\right)}\right]\Biggr\}\notag\\
&\quad+\dfrac{m_k^4}{8E_0^3}\Biggl\{\dfrac{Mb}{1+\Lambda^2}\left[\dfrac{1}{\sqrt{r_D^2-b^2\left(1+\Lambda^2\right)}}+\dfrac{1}{\sqrt{r_S^2-b^2\left(1+\Lambda^2\right)}}\right]\notag\\
&\quad+\dfrac{1}{\sqrt{1+\Lambda^2}}\left[\sqrt{r_D^2-b^2\left(1+\Lambda^2\right)}+\sqrt{r_S^2-b^2\left(1+\Lambda^2\right)}\right]\notag\\
&\quad+\dfrac{M}{\left(1+\Lambda^2\right)^{\frac{3}{2}}}\left(\sqrt{\dfrac{r_D-b\sqrt{1+\Lambda^2}}{r_D+b\sqrt{1+\Lambda^2}}}+\sqrt{\dfrac{r_S-b\sqrt{1+\Lambda^2}}{r_S+b\sqrt{1+\Lambda^2}}}\right)\notag\\
&\quad-\dfrac{M\left(2+\Lambda^2\right)}{\left(1+\Lambda^2\right)^{\frac{3}{2}}}\ln\left[\dfrac{\left(r_D-\sqrt{r_D^2-b^2\left(1+\Lambda^2\right)}\right)\left(r_S-\sqrt{r_S^2-b^2\left(1+\Lambda^2\right)}\right)}{b^2\left(1+\Lambda^2\right)}\right]\Biggr\}\notag\\
&= \Phi_{k1}+\Phi_{k2}.
\end{align}
\end{widetext}
The phase $\Phi_{k,1}$ of the expansion to order $\mathcal{O}\left(\frac{m_k^2}{E_0^2}\right)$ is represented by coefficient $\frac{m_k^2}{E_0^2}$, while the phase $\Phi_{k,2}$ to order $\mathcal{O}\left(\frac{m_k^4}{E_0^4}\right)$ is represented by the remainder. By expanding the above equation (\ref{PhiS0D}) above up to $\frac{b^2}{r_{S,D}^2}$ for $b\ll r_{S,D}$, we find
\begin{align}
\Phi_{k,1}
&\simeq \dfrac{m_k^2}{2E_0}\biggl\{\sqrt{1+\Lambda^2}\left(r_D+r_S\right)\notag\\
&\quad\times \left(1-\dfrac{1+\Lambda^2}{2}\dfrac{b^2}{r_Dr_S}+\dfrac{2M}{r_D+r_S}\right)\notag\\
&\quad\times \dfrac{M\Lambda^2}{\sqrt{1+\Lambda^2}}\ln\left[\dfrac{4r_Dr_S}{b^2\left(1+\Lambda^2\right)}\right]\biggr\}
\end{align}
and
\begin{align}
\Phi_{k,2}
&\simeq \dfrac{m_k^4}{2E_0^3}\biggl\{\dfrac{r_D+r_S}{\sqrt{1+\Lambda^2}}\notag\\
&\quad\times \left(1-\dfrac{1+\Lambda^2}{2}\dfrac{b^2}{r_Dr_S}+\dfrac{1}{1+\Lambda^2}\dfrac{2M}{r_D+r_S}\right)\notag\\
&\quad\times \dfrac{M\left(2+\Lambda^2\right)}{\left(1+\Lambda^2\right)^{\frac{3}{2}}}\ln\left[\dfrac{4r_Dr_S}{b^2\left(1+\Lambda^2\right)}\right]\biggr\}.
\end{align}
It should be pointed out that the gravitational lensing of neutrinos appear in the process of propagation. In order to scrutinize the neutrino flavour oscillation probability (\ref{PhiS0D}) around the black-hole-like wormhole, we must derive the phase difference along the different paths \cite{cardall1997neutrino, fornengo1997gravitational},
\begin{align}
\label{deltaPhi1}
\Delta\Phi_{ij,1}^{pq}
&= \Phi_{i,1}^{p}-\Phi_{j,1}^{q}\notag\\
&=\Delta m_{ij}^2 A_{pq,1}+\Delta b_{pq}^2 B_{ij,1}+C_{ij,1}^{pq},
\end{align}
where
\begin{align}
\label{deltam1}
\Delta m_{ij}^2 &= m_i^2-m_j^2,\\
\Delta b_{pq}^2 &= b_p^2-b_q^2,\\
\label{Apq1}
A_{pq,1} &= \sqrt{1+\Lambda^2}\dfrac{r_S+r_D}{2E_0}\biggl(1+\dfrac{2M}{r_S+r_D}\notag\\
&\quad -\dfrac{1+\Lambda^2}{4}\dfrac{\sum b_{pq}^2}{r_Sr_D}\biggl),\\
\label{Bij1}
B_{ij,1} &= -\left(1+\Lambda^2\right)^{\frac{3}{2}}\dfrac{\sum m_{ij}^2}{8E_0}\left(\dfrac{1}{r_S}+\dfrac{1}{r_D}\right),\\
\label{Cij1pq}
C_{ij,1}^{pq} &= \dfrac{M\Lambda^2}{2E_0\sqrt{1+\Lambda^2}}\ln\left[\dfrac{\left(b_{q}^{2}\right)^{m_j^2}}{\left(b_{p}^{2}\right)^{m_i^2}}\left(\dfrac{4r_Dr_S}{1+\Lambda^2}\right)^{m_i^2-m_j^2}\right],\\
\label{bpq}
\sum b_{pq}^2 &= b_p^2+b_q^2,\\
\sum m_{ij}^2 &= m_i^2+m_j^2,
\end{align}
and
\begin{align}
\label{deltaPhi2}
\Delta\Phi_{ij,2}^{pq}
&= \Phi_{i,2}^{p}-\Phi_{j,2}^{q}\notag\\
&=\Delta m_{ij}^4 A_{pq,2}+\Delta b_{pq}^2 B_{ij,2}+C_{ij,2}^{pq},
\end{align}
where
\begin{align}
\label{deltam2}
\Delta m_{ij}^4 &= m_i^4-m_j^4,\\
\Delta b_{pq}^2 &= b_p^2-b_q^2,\\
\label{Apq2}
A_{pq,2} &= \dfrac{1}{\sqrt{1+\Lambda^2}}\dfrac{r_S+r_D}{8E_0^3}\biggl(1+\dfrac{1}{1+\Lambda^2}\dfrac{2M}{r_S+r_D}\notag\\
&\quad -\dfrac{1+\Lambda^2}{4}\dfrac{\sum b_{pq}^2}{r_Sr_D}\biggl),\\
\label{Bij2}
B_{ij,2} &= -\sqrt{1+\Lambda^2}\dfrac{\sum m_{ij}^2}{32E_0^3}\left(\dfrac{1}{r_S}+\dfrac{1}{r_D}\right),\\
\label{Cij2pq}
C_{ij,2}^{pq} &= \dfrac{M\Lambda^2}{8E_0^3\left(1+\Lambda^2\right)^{\frac{3}{2}}}\ln\left[\dfrac{\left(b_{q}^{2}\right)^{m_j^4}}{\left(b_{p}^{2}\right)^{m_i^4}}\left(\dfrac{4r_Dr_S}{1+\Lambda^2}\right)^{m_i^4-m_j^4}\right].
\end{align}
Combining with the Eqs.(\ref{deltaPhi1})-(\ref{Cij2pq}), we get the whole phase difference,
\begin{align}
\label{deltaPhi}
\Delta\Phi_{ij}^{pq} &=\Delta\Phi_{ij,1}^{pq}+\Delta\Phi_{ij,2}^{pq}\notag\\
&= \Delta m_{ij}^2\mathcal{A}_{ij}^{pq}+\Delta b_{pq}^2\mathcal{B}_{ij}+\mathcal{C}_{ij}^{pq},
\end{align}
where
\begin{align}
\mathcal{A}_{ij}^{pq}&= A_{pq,1}+\sum m_{ij}^2A_{pq,2},\\
\mathcal{B}_{ij}&= B_{ij,1}+B_{ij,2},\\
\mathcal{C}_{ij}^{pq}&= C_{ij,1}^{pq}+C_{ij,2}^{pq}.
\end{align}
Although the local energy of the neutrinos going through the black-hole-like wormhole is corrected with the factor $\Lambda^{2}$, the wormhole shift affects most terms of the expression for neutrino lensing probability. It is fundamental that the Damour-Solodukhin parameter generates four terms forming the two parts respectively in the phase difference above. The phase corresponding to the routes should be added the upper indices like $\Phi_{i}^{p}$ to mark the trajectory that the neutrinos travel along and the trajectories have their impact factor $b_{p}$. According to the equations above, the phase difference of the transition probability for propagating neutrinos along the various paths through the Damour-Solodukhin wormhole relates to the individual masses of neutrinos denoted as $m_{i}$, the difference between the squared neutrino masses like $\Delta m_{ij}^{2}$ or $\Delta m_{ij}^{4}$ and the structure of this kind of gravitational source. The phase difference (\ref{deltaPhi}) contains several terms specified by the Eqs.(\ref{deltaPhi1})-(\ref{Cij2pq}). It is clear that one term in the bracket of the Eq.(\ref{deltaPhi}) is the product of $\Delta m_{ij}^2$, $\Delta m_{ij}^4$ and $\mathcal{A}_{pq}$. When $\Lambda=0$, the phase difference will recover to be the results in \cite{swami2020signature}. The coefficient $\mathcal{B}_{ij}$ is independent of the Damour-Solodukhin factor $\Lambda$, but depends on the neutrino masses. The coefficient $\mathcal{C}_{ij}^{pq}$ under the influence from $\Lambda$ is also a function of the individual neutrino masses labeled as $m_{i}$. The coefficients $\mathcal{A}_{pq}$ and $\mathcal{B}_{ij}$ are invariant by exchanging between their own lower indices. Based on the exchange $p\leftrightarrow q$ and $i\leftrightarrow j$, the sign of $\mathcal{C}_{ij}^{pq}$ changes.

\section{Gravitational lensing of neutrinos in Damour-Solodukhin spacetime}

In the background of gravitational source, the neutrinos may propagate nonradially and the gravitational lensing can appear between the neutrino emitter and the receiver \cite{cardall1997neutrino}. The propagation of neutrinos around the spherically symmetric black-hole-like wormhole can be shown in the Fig.\ref{fig:diagram}, which is similar to that in Ref.\cite{cardall1997neutrino}. The gravitational lensing prefers that the neutrinos traveling through different paths meet the detector $D$, so the neutrino flavour eigenstate must be rewritten as \cite{chetouani1984geometrical,crocker2004neutrino,maki1962remarks,pontecorvo1967neutrino,stodolsky1979matter,swami2021aspects},
\begin{align}
\ket{\nu_{\alpha}(t_{D},\textbf{x}_{D})}=N\sum_{i}U_{\alpha i}^{\ast}
\sum_{p}\exp\left(-i\Phi_{i}^{p}\right)\ket{\nu_{i}(t_{S},\textbf{x}_{S})},
\end{align}
where $p$ represents the path index. Based on that nearly all neutrinos focus on the detector, the oscillation probability for $\nu_{\alpha}\rightarrow\nu_{\beta}$ at the detection point is given by \cite{chetouani1984geometrical,crocker2004neutrino,maki1962remarks,pontecorvo1967neutrino,stodolsky1979matter,swami2021aspects},
\begin{align}
\label{Philens}
\mathcal{P}_{\alpha\beta}^{\mathrm{lens}}&=|\langle\nu_{\beta}|\nu_{\alpha}(t_{D}, \textbf{x}_D)\rangle|^{2}\hspace{2cm}\notag\\
&=|N|^{2}\sum_{i, j}U_{\beta i}U_{\beta j}^{\ast}U_{\alpha j}U_{\alpha j}^{\ast}\sum_{p, q}\exp(\Delta\Phi_{ij}^{pq}),
\end{align} 
and the normalization constant is \cite{cardall1997neutrino},
\begin{align}
|N|^{2}=\left[\sum_{i}|U_{\alpha i}|^{2}\sum_{p,q}\exp(-i\Delta\Phi_{ij}^{pq})\right]^{-1}.
\end{align}
In view of our discussion on the phase difference $\Delta\Phi_{ij}^{pq}$ in the expressions, there are also effects from the individual masses and difference between the squared masses of neutrinos, the wormhole structure on the likelihood of neutrino oscillation in the case of neutrino lensing from Eq.(\ref{Philens}), which is similar to those around the spherically symmetric sources such as Schwarzschild black hole Ref.\cite{swami2020signature}. 

We narrow our consideration on the neutrino oscillation probability for gravitational lensing of neutrinos to elucidate the influence of lens with the factor $\Lambda$. In the background of Damour-Solodukhin wormhole as lens for two-flavour neutrinos, we probe the flavour transition probabilities for $\nu_{\alpha}\to\nu_{\beta}$ at the detection position. We derive the probability function under the weak field limit in the plane located by the source, lens and detector \cite{swami2020signature, cardall1997neutrino, fornengo1997gravitational, swami2021aspects},
\begin{widetext}
\begin{align}
\label{Pab2}
\mathcal{P}_{\alpha\beta}^{\mathrm{lens}}
&= \left|N\right|^2\biggl\{2\sum_i\left|U_{\beta i}\right|^2\left|U_{\alpha i}\right|^2\left[1+\cos\left(\Delta b_{12}^2\mathcal{B}_{ii}+\mathcal{C}_{ii}^{12}\right)\right]\notag\\
&\quad+\sum_{i\neq j}U_{\beta i}U_{\beta j}^*U_{\alpha j}U_{\alpha i}^*\left[\exp\left(-\mathrm{i}\Delta m_{ij}^2 \mathcal{D}_{ij}^{11}\right)+\exp\left(-\mathrm{i}\Delta m_{ij}^2 \mathcal{D}_{ij}^{22}\right)\right]\notag\\
&\quad+\sum_{i\neq j}U_{\beta i}U_{\beta j}^*U_{\alpha j}U_{\alpha i}^*\left[\cos\left(\Delta b_{12}^2\mathcal{B}_{ij}+\mathcal{C}_{ij}^{12}\right)-\mathrm{i}\sin\left(\Delta b_{12}^2\mathcal{B}_{ij}+\mathcal{C}_{ij}^{12}\right)\right]\exp\left(-\mathrm{i}\Delta m_{ij}^2\mathcal{A}_{ij}^{12}\right)\notag\\
&\quad+\sum_{i\neq j}U_{\beta i}U_{\beta j}^*U_{\alpha j}U_{\alpha i}^*\left[\cos\left(\Delta b_{21}^2\mathcal{B}_{ij}+\mathcal{C}_{ij}^{21}\right)-\mathrm{i}\sin\left(\Delta b_{21}^2\mathcal{B}_{ij}+\mathcal{C}_{ij}^{21}\right)\right]\exp\left(-\mathrm{i}\Delta m_{ij}^2\mathcal{A}_{ij}^{21}\right)\biggr\},
\end{align}
\end{widetext}
where
\begin{align}
\Delta m_{ij}^{2}\mathcal{A}_{ij}^{pp}+C_{ij}^{pp}:=\Delta m_{ij}^2\mathcal{D}_{ij}^{pp}.
\end{align}
It is necessary to explain the terms enclosed in curly brakets in the probability expression (\ref{Pab2}). The first term stands for the situation with $i=j$. The second one represents the case with $i\neq j$ but $p=q$. The third and fourth term both relate to $i\neq j$ and $p\neq q$, one for $p<q$ and the other for $p>q$. In the case of two flavours for the neutrinos, the leptonic mixing matrix may be shown as the $2\times2$ ones with mixing angle $\alpha$ \cite{esteban2019global},
\begin{align}
\label{U}
U\equiv\left(\begin{matrix}
\cos\alpha&\sin\alpha\\
-\sin\alpha&\cos\alpha
\end{matrix}\right).
\end{align}
By substituting the mixing matrix (\ref{U}) into the Eq.(\ref{Pab2}), the probability of oscillation $\nu_{e}\to\nu_{\mu}$ around the Damour-Solodukhin wormhole is obtained,
\begin{widetext}
\begin{align}
\label{Pab_lens}
\mathcal{P}_{\alpha\beta}^{\mathrm{lens}}
&=\left|N\right|^2\sin^22\alpha\biggl\{\sin^2\left(\dfrac{\Delta m_{12}^2\mathcal{A}_{12}^{11}+\mathcal{C}_{12}^{11}}{2}\right)+\sin^2\left(\dfrac{\Delta m_{12}^2\mathcal{A}_{12}^{22}+\mathcal{C}_{12}^{22}}{2}\right)\notag\\
&\quad+\dfrac{1}{2}\cos\left(\Delta b_{12}^2\mathcal{B}_{11}+\mathcal{C}_{11}^{12}\right)+\dfrac{1}{2}\cos\left(\Delta b_{12}^2\mathcal{B}_{22}+\mathcal{C}_{22}^{12}\right)\notag\\
&\quad-\dfrac{1}{2}\cos\left(\Delta m_{12}^2\mathcal{A}_{12}\right)
\left[\cos\left(\Delta b_{12}^2\mathcal{B}_{12}+\mathcal{C}_{12}^{12}\right)+\cos\left(\Delta b_{21}^2\mathcal{B}_{12}+\mathcal{C}_{12}^{21}\right)\right]\notag\\
&\quad+\dfrac{1}{2}\sin\left(\Delta m_{12}^2\mathcal{A}_{12}\right)
\left[\sin\left(\Delta b_{12}^2\mathcal{B}_{12}+\mathcal{C}_{12}^{12}\right)+\sin\left(\Delta b_{21}^2\mathcal{B}_{12}+\mathcal{C}_{12}^{21}\right)\right]\biggr\}.
\end{align}
\end{widetext}
In view of the leptonic mixing matrix (\ref{U}) and the phase difference \ref{deltaPhi} through various trajectories of neutrinos, the Eq.(25) displays the normalization constant,
\begin{align}
\left|N\right|^2
&=\biggl[2+2\cos^2\alpha\cos\left(\Delta b_{12}^2\mathcal{B}_{11}+\mathcal{C}_{11}^{12}\right)\notag\\
&\quad+2\sin^2\alpha\cos\left(\Delta b_{12}^2\mathcal{B}_{22}+\mathcal{C}_{22}^{12}\right)\biggl]^{-1}.
\end{align}
The wormhole factor $\Lambda$ also modifies these quantities. It is obvious that $\mathcal{C}_{ij}^{pq}$ disappears when $\Lambda=0$, which lead the probability and  the normalization constant a of two neutrinos' flavor oscillation to return to be those of the Schwarzschild black hole \cite{swami2020signature}.

\section{Numerical Results}

\begin{figure}
\centering
\includegraphics[width=8cm]{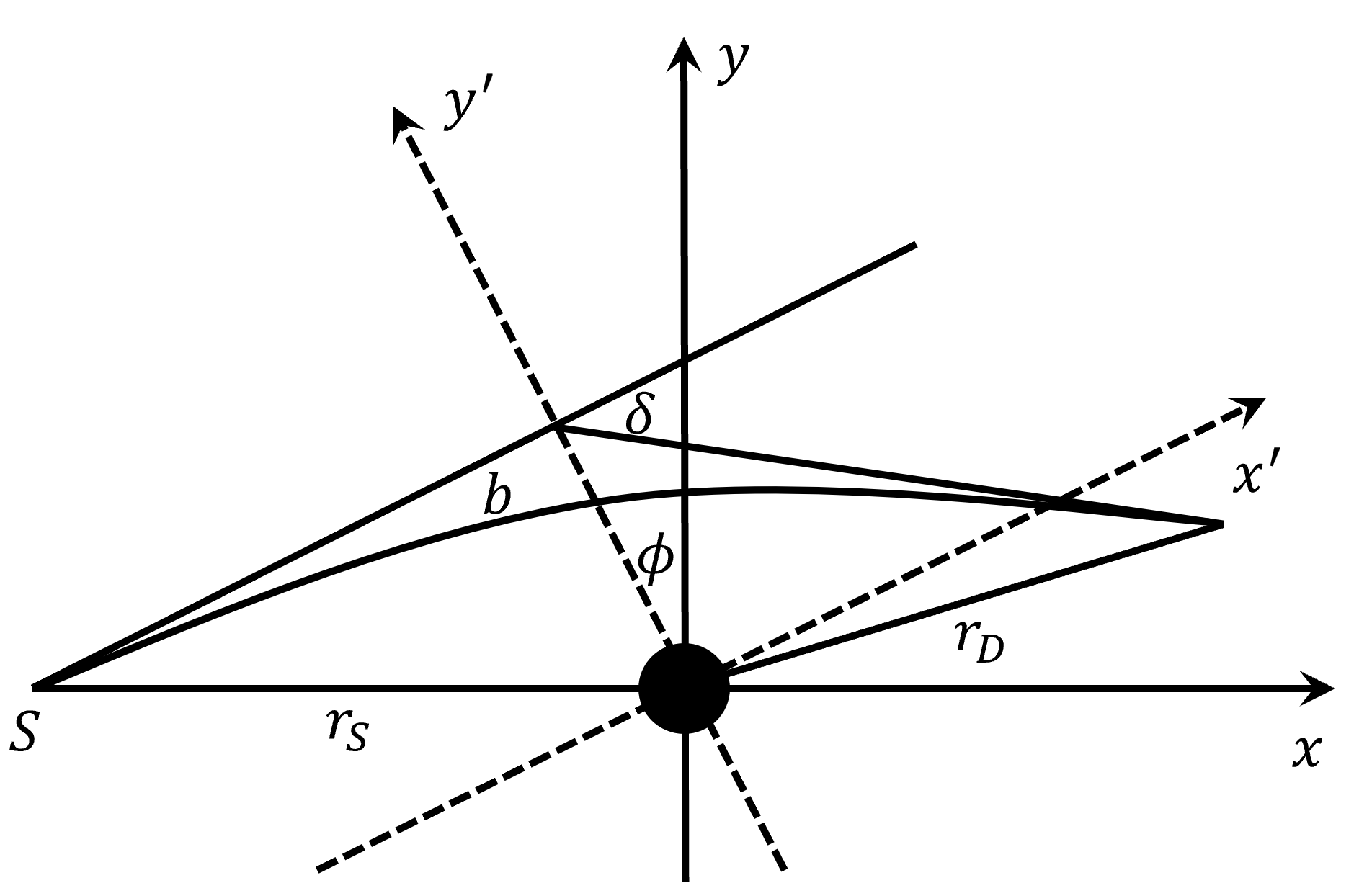}
\caption{\label{fig:diagram} Diagrammatic illustration of the Damour-Solodukhin spacetime with mild lensing of neutrinos. The metric describes the propagation of neutrinos from the source ($S$) to detector ($D$) in the exterior of a static, non-spherical large object.}
\end{figure}

For the sake of elucidating the neutrino oscillation around the Damour-Solodukhin source quantitatively, it is necessary to acquire a better understanding of the lensing probabilities from Eq.(\ref{Pab_lens}). The Fig.\ref{fig:diagram} illustrates that a neutrino source ($S$), the shifted Schwarzschild black hole as a gravitational source and a plane-based detector ($D$) comprise a system \cite{damour2007wormholes,swami2020signature}. The wormhole is located at the coordinate system's origin in the $(x,y)$ coordinate system, and the physical distances $r_S$ and $r_D$ separate the neutrino source from the lens and detector respectively. It is also feasible to think of the coordinate system $(x',y')$ as the coordinate system $(x,y)$ with rotation angle $\phi$, with $y'=-x\sin\phi+y\cos\phi$ and $x'=x\cos\phi+y\sin\phi$ \cite{swami2020signature}. Remarkably, the three parts of the apparatus are co-linear in the plane when $\phi=0$. Referring to Ref.\cite{swami2020signature}, the relationship between the impact parameter $b$ and $\delta$, the deflection angle of the neutrino from its initial route due to the Damour-Solodukhin wormhole is as follows,
\begin{align}
\label{delta}
\delta\sim\dfrac{y_D'-b}{x_D'}=-\dfrac{4M}{b\left(1+\Lambda^2\right)}=-\dfrac{2r_H}{b},
\end{align}
where the detector's location is $(x_D',y_D')$ and $r_H=\left(2+\Lambda^2\right)M$ is the event horizon radius. Now that we have the identity $\sin\phi = \frac{b}{r_S}$ from Fig.\ref{fig:diagram}, we can rewrite the Eq.(\ref{delta}) as
\begin{align}
\label{solve_b}
\left(2r_Hx_D+by_D\right)\sqrt{1-\dfrac{b^2}{r_S^2}}=b^2\left(\dfrac{x_D}{r_S}+1\right)-\dfrac{2r_Hby_D}{r_S}.
\end{align}

We are going to compute the wormhole lensing probability for neutrino oscillation to reveal the $\Lambda$-term effect. It is worth comparing our results with those of Schwarzschild black hole from Ref.\cite{swami2020signature} because the Damour-Solodukhin wormhole is just a corrected Schwarzschild black hole \cite{damour2007wormholes}. The impact parameters such as $r_S$, $r_H$, and the lensing position $(x_D,y_D)$ can be obtained from Eq.(\ref{delta}). In order to show the influence of the deviation on the conversion probabilities distinctly and the trend of changes clearly, we choose the values of $\Lambda$ to be rather large in the figures below.

\subsection{Two flavor case}

\begin{figure*}
\centering
\includegraphics[height=5cm]{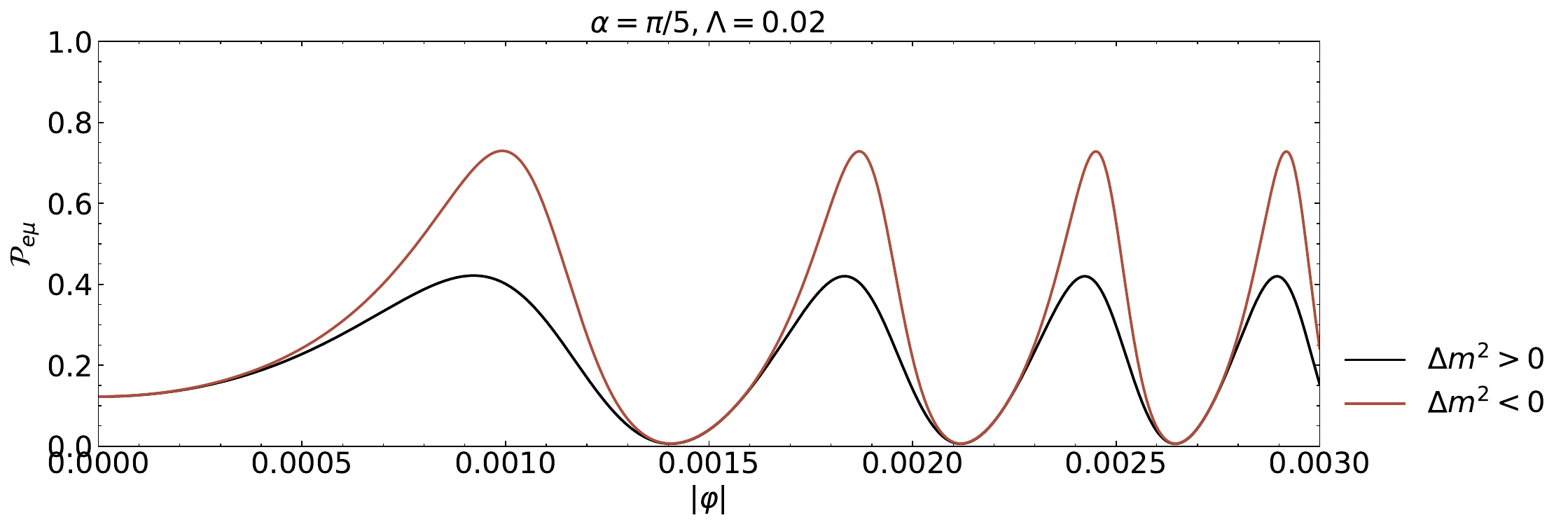}
\includegraphics[height=5cm]{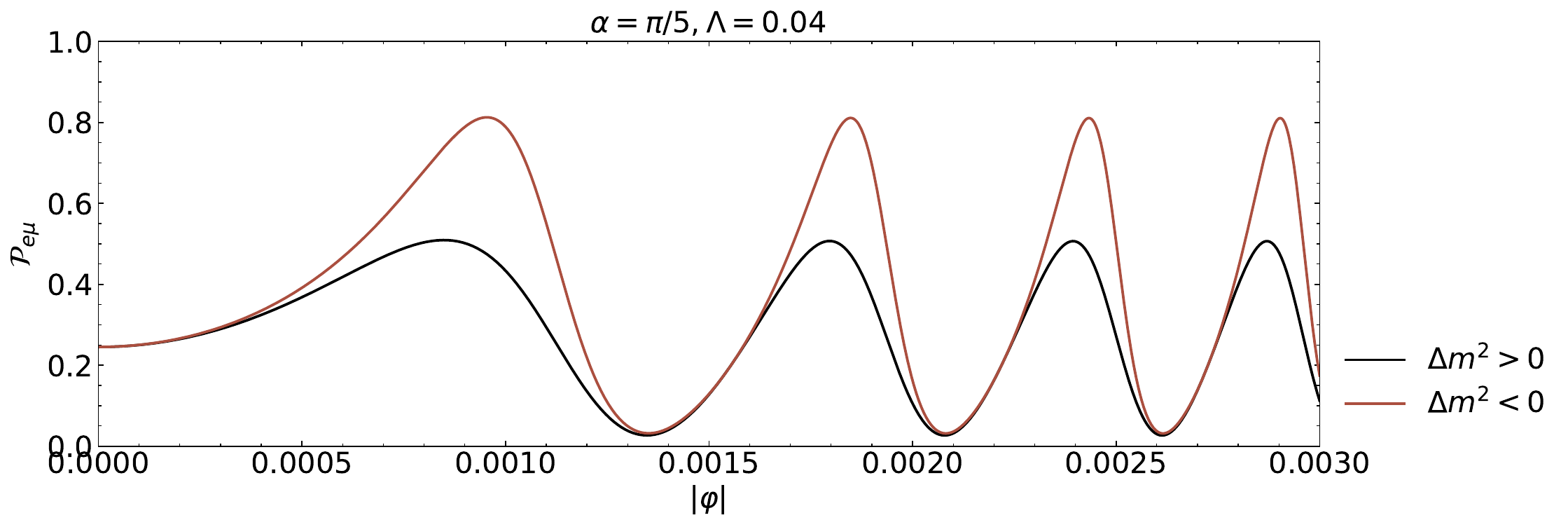}
\includegraphics[height=5cm]{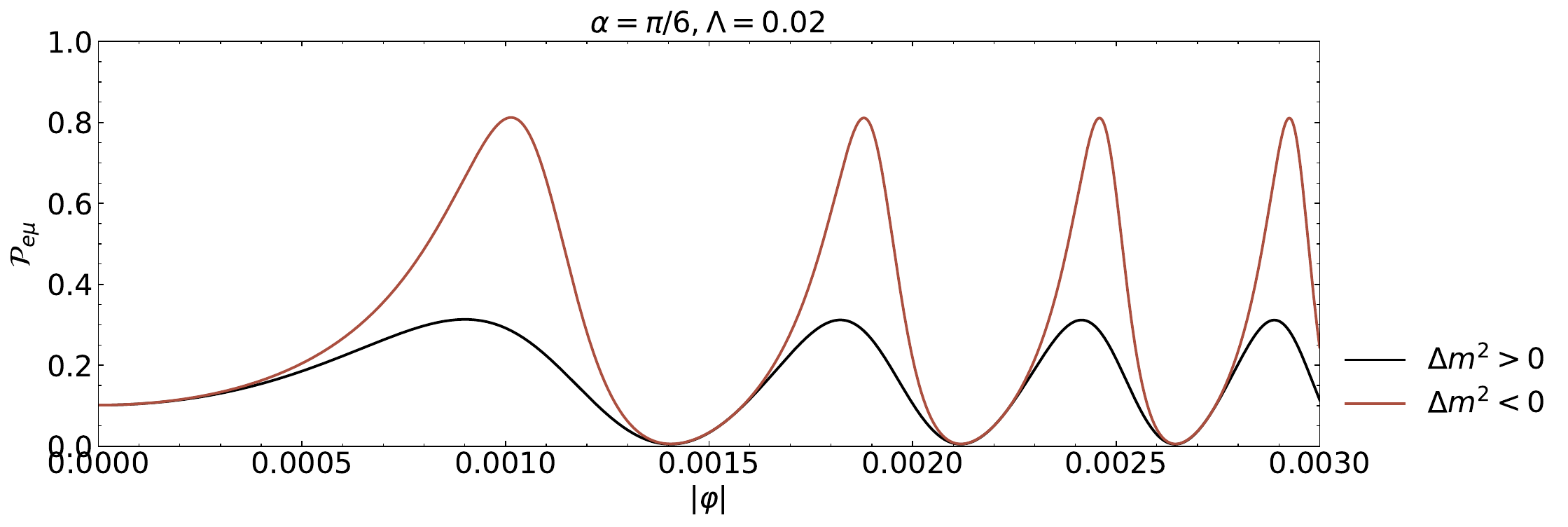}
\includegraphics[height=5cm]{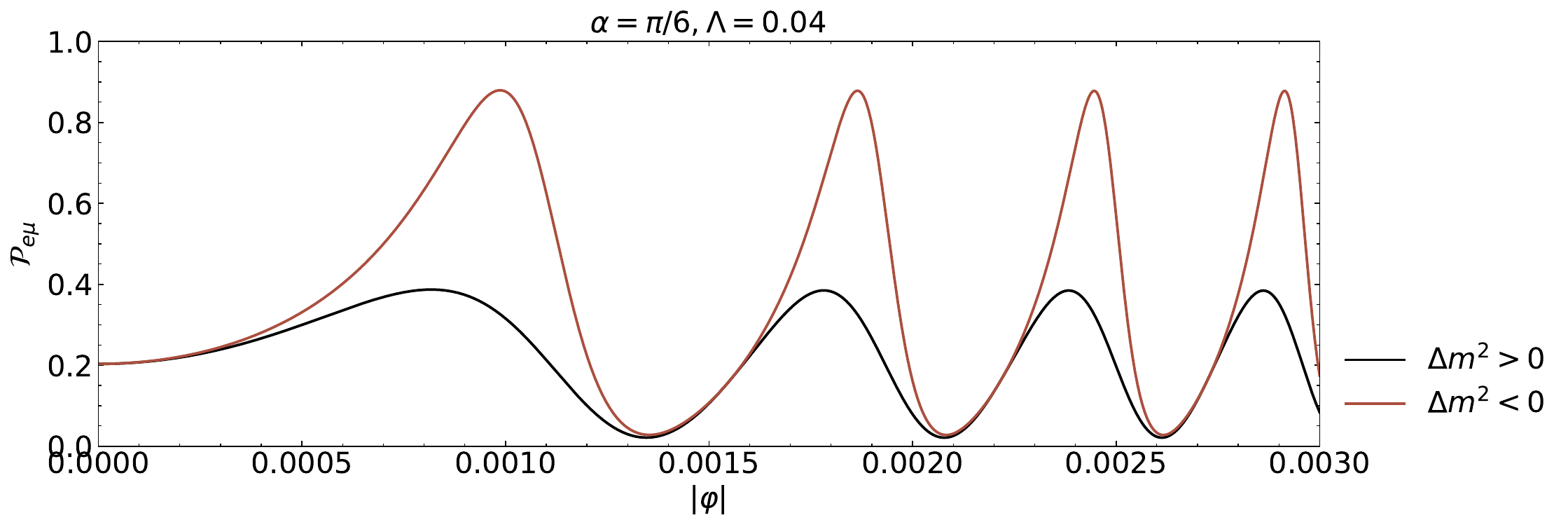}
\caption{\label{fig:prob1} The probability for $\Lambda=0.02, 0.04$ of $\nu_e\to \nu_{\mu}$ conversion based on azimuthal angle $\phi$ is examined for both normal and inverted neutrino mass orderings in the two flavor scenario when the mixing angles are $\alpha=\frac{\pi}{5}$ and $\alpha=\frac{\pi}{6}$. Here, $r_D=10^8\,\mathrm{km}$, $r_S=10^5r_D$, $E_0=10\,\mathrm{MeV}$, and $|\Delta m^2|=10^{-3}\,\mathrm{eV^2}$. It is assumed that the lightest neutrino is massless.}
\end{figure*}

\begin{figure*}
\centering
\includegraphics[height=5cm]{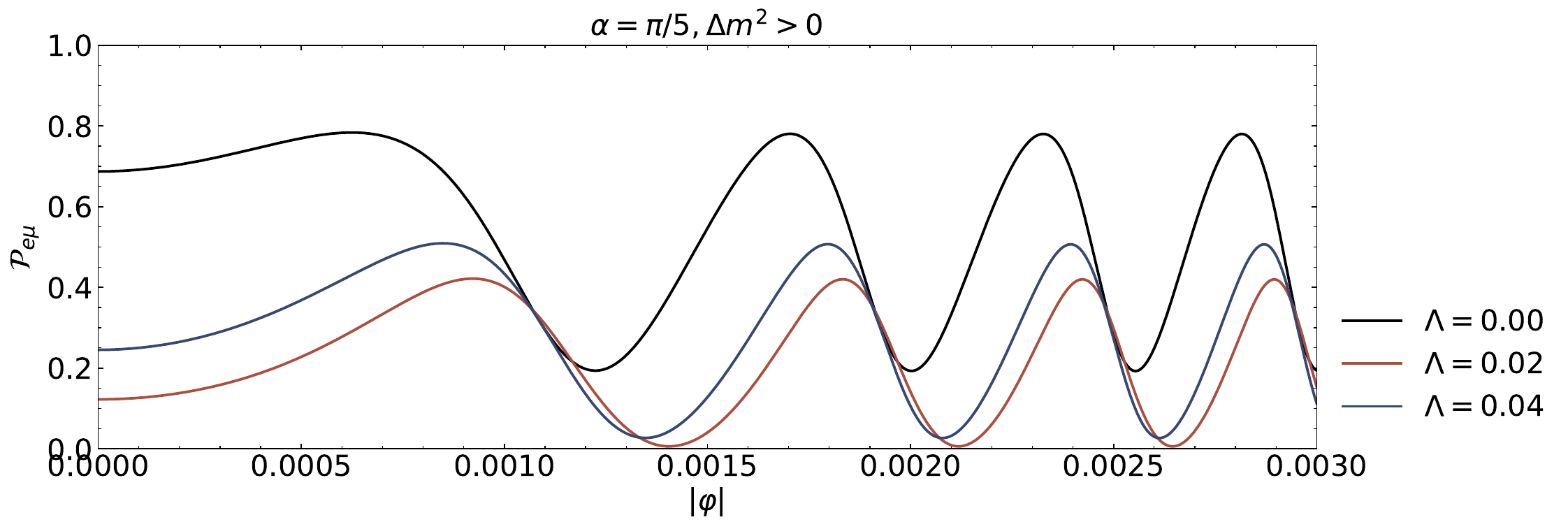}
\includegraphics[height=5cm]{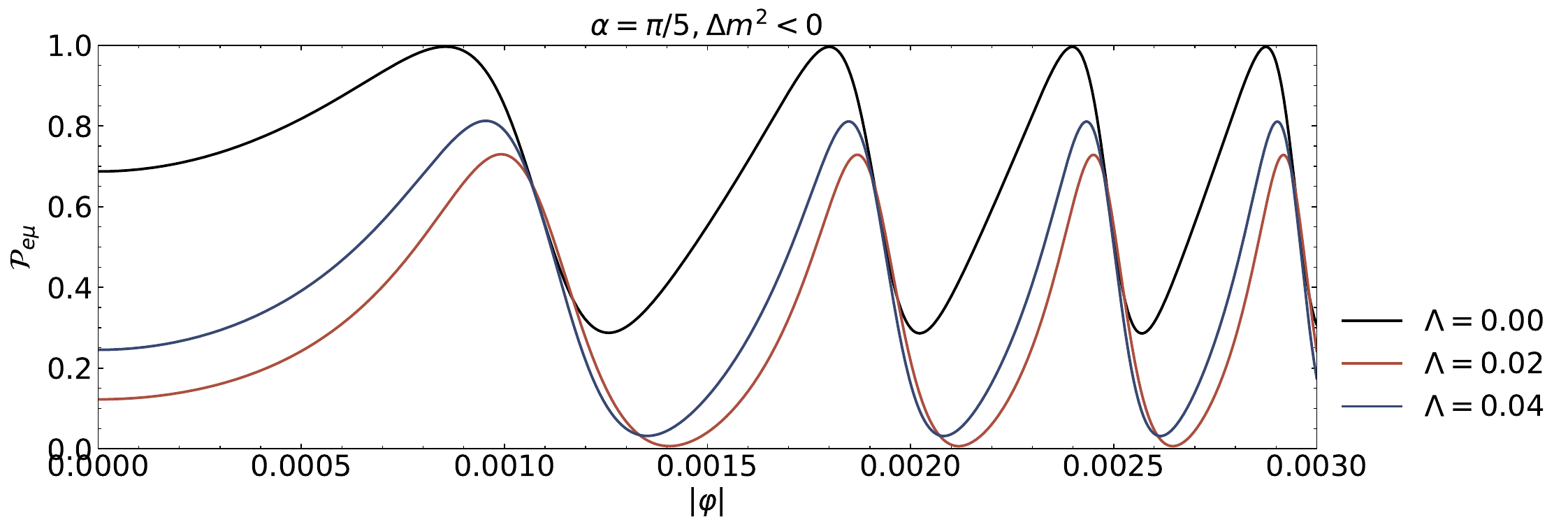}
\includegraphics[height=5cm]{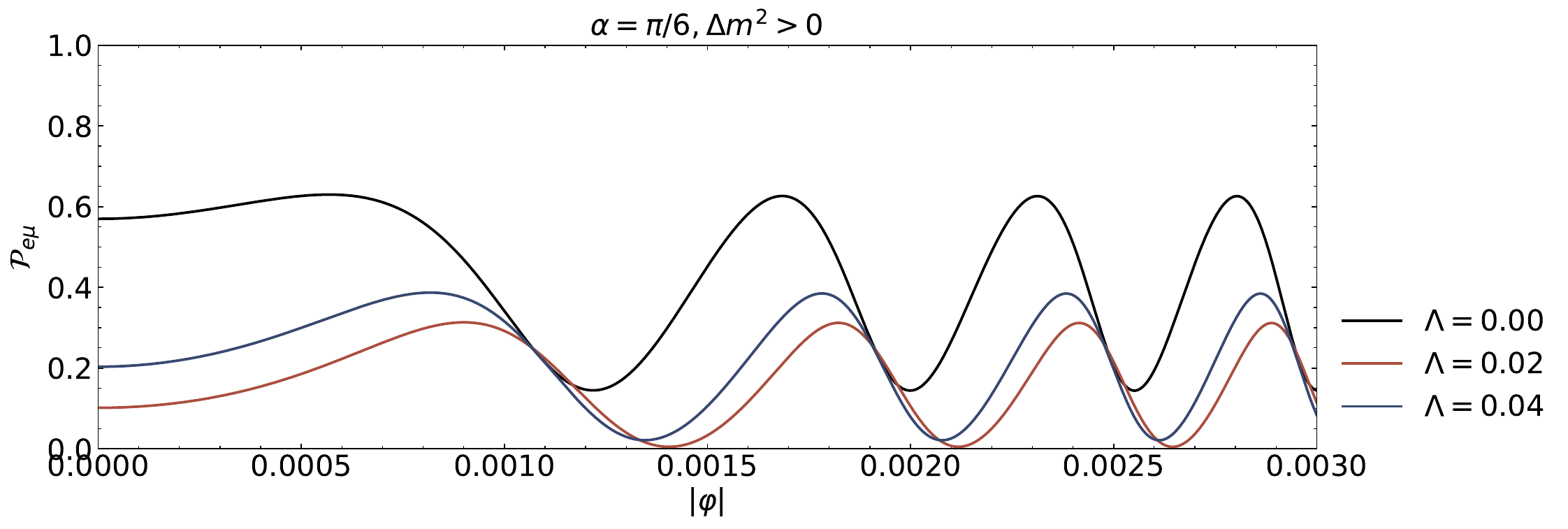}
\includegraphics[height=5cm]{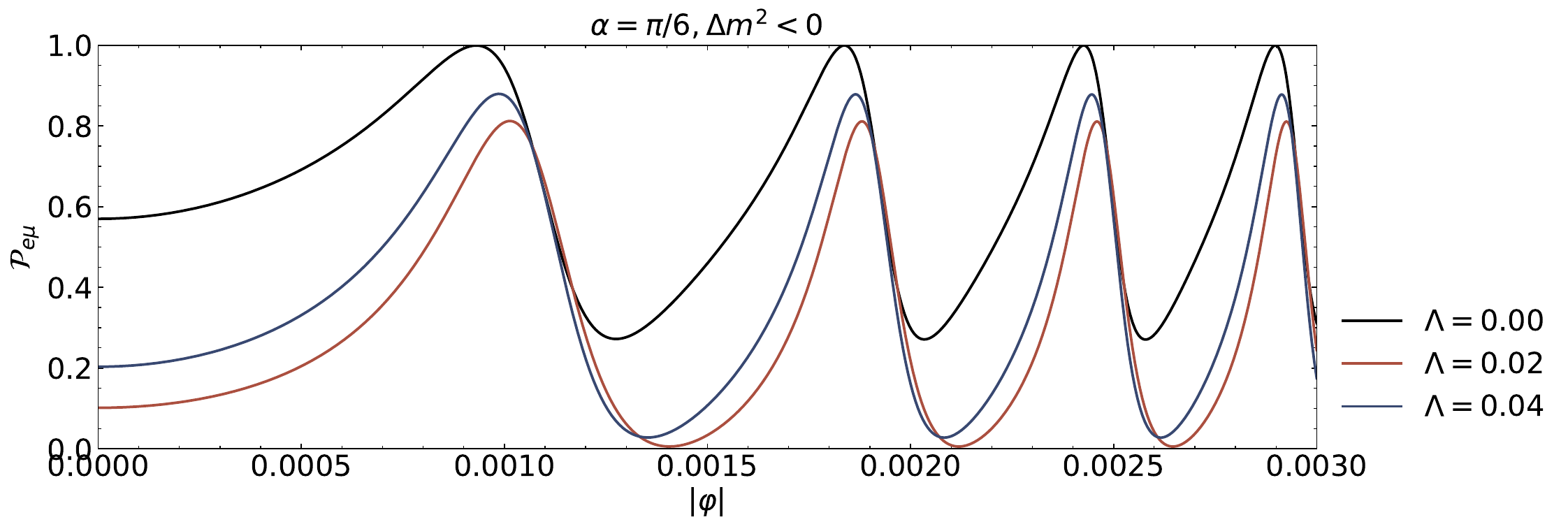}
\caption{\label{fig:prob2} The probability of $\nu_e\to \nu_{\mu}$ conversion based on azimuthal angle $\phi$ for the mixing angles $\alpha=\frac{\pi}{5}$ and $\alpha=\frac{\pi}{6}$ when $\Lambda=0, 0.02, 0.04$ . Here, $r_D=10^8\,\mathrm{km}$, $r_S=10^5r_D$, $E_0=10\,\mathrm{MeV}$ and $|\Delta m^2|=10^{-3}\,\mathrm{eV^2}$, and the lightest neutrino is considered to be massless.}
\end{figure*}

\begin{figure*}
\centering
\includegraphics[height=5cm]{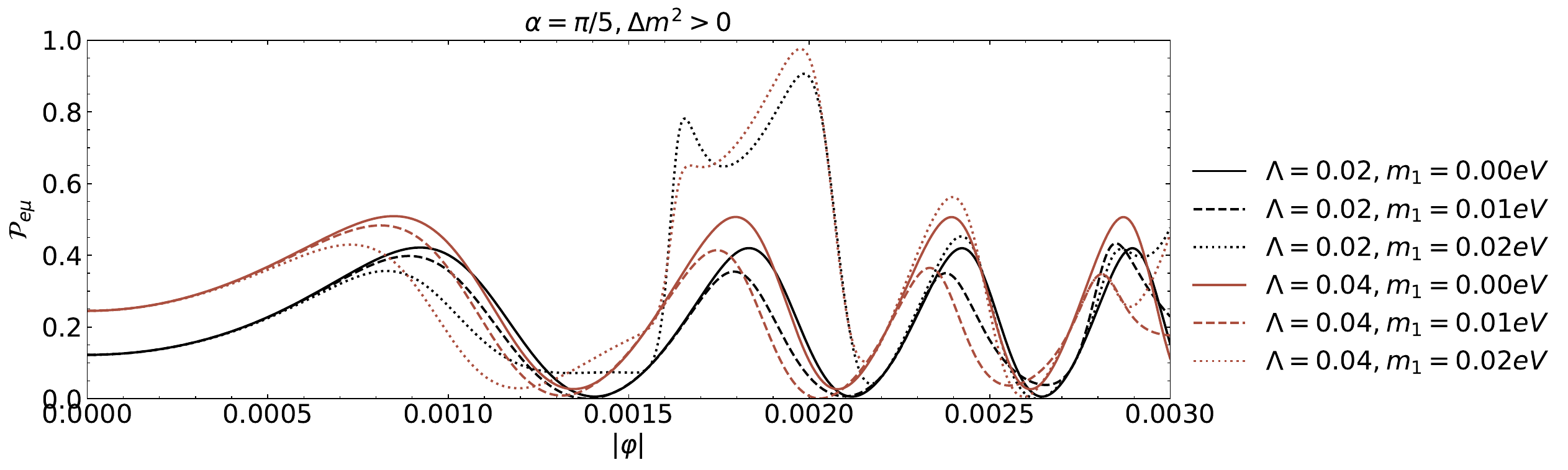}
\includegraphics[height=5cm]{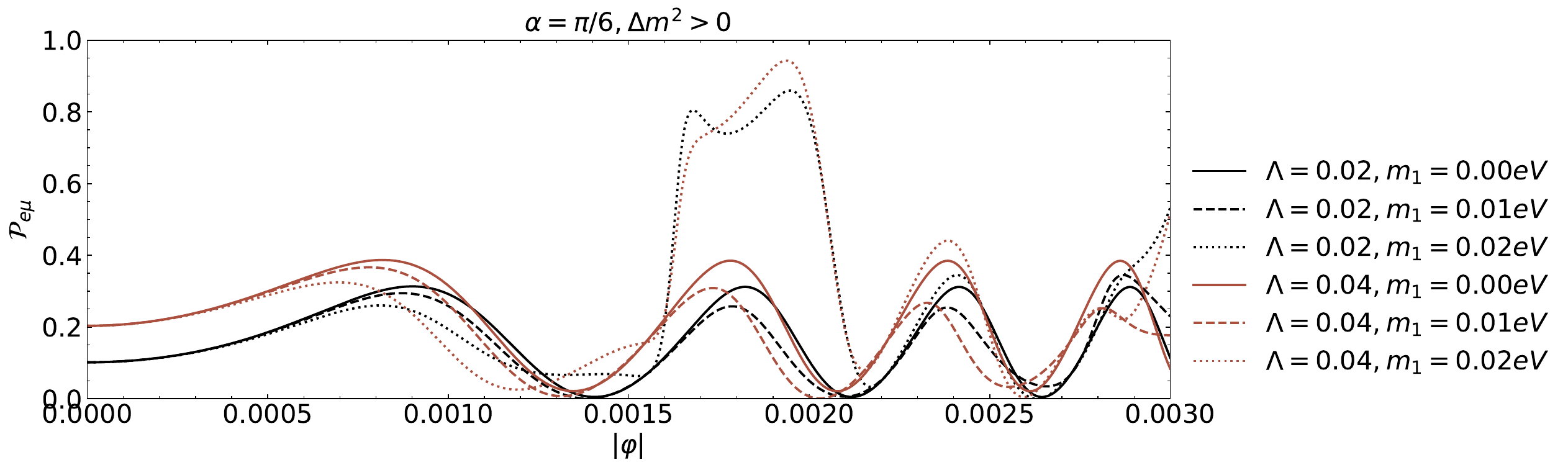}
\includegraphics[height=5cm]{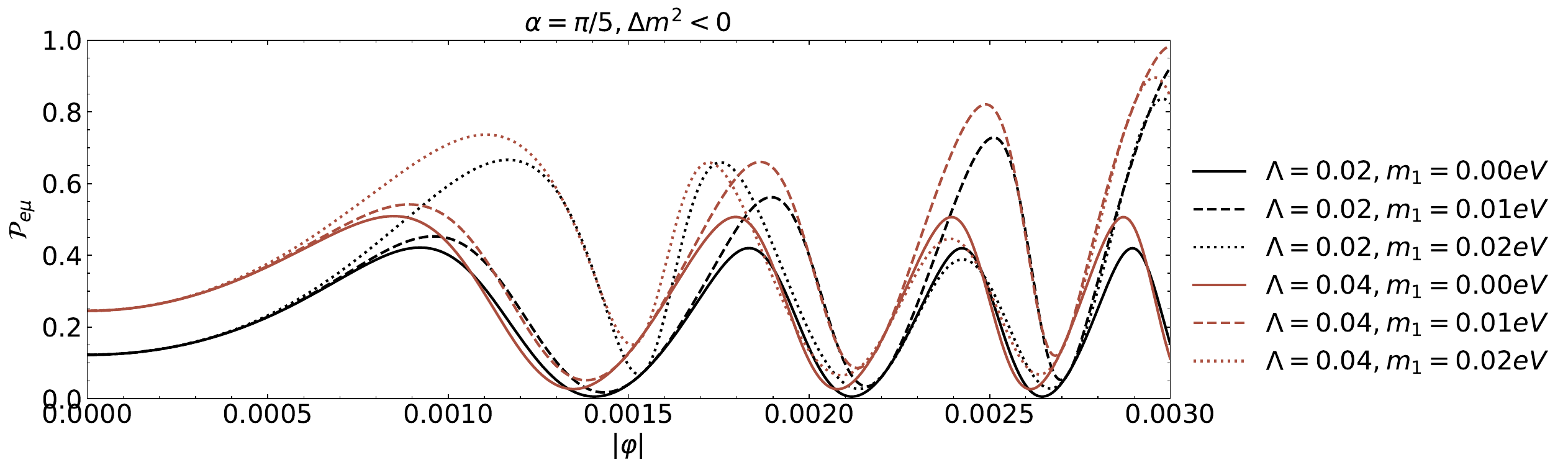}
\includegraphics[height=5cm]{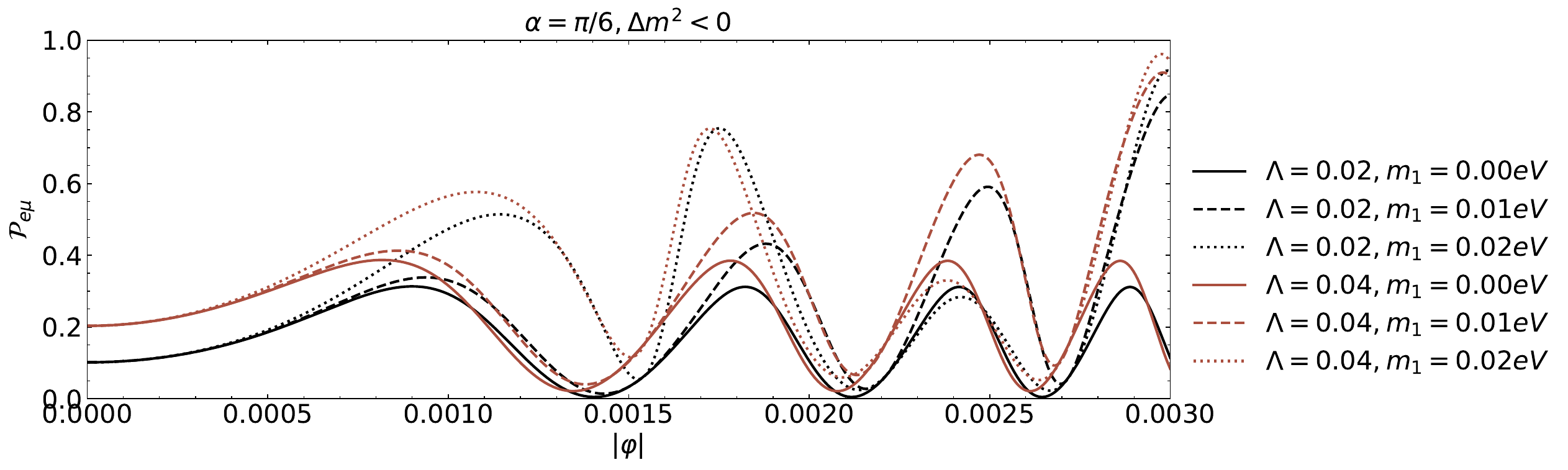}
\caption{\label{fig:prob3} The first and second diagrams: the probability of neutrino oscillation vs azimuthal angle $\phi$ for $\Lambda=0.02$ (black lines) and $0.04$ (red lines), and $\Delta m^2>0$ (normal ordering). Where $m_1=0\,\mathrm{eV}$ is shown by the solid line, $m_1=0.01\,\mathrm{eV}$ is shown by the dashed line, and $m_1=0.02\,\mathrm{eV}$ is indicated by the dotted line. The third and forth planes display the probability of neutrino oscillations against azimuthal angle $\phi$ (inverted ordering) for values of $\Lambda = 0.02$ (black lines) and $0.04$ (red lines), and $\Delta m^2<0$. The solid line indicates $m_1=0\,\mathrm{eV}$, dashed line shows $m_1=0.01\,\mathrm{eV}$ and the dotted line suggests $m_1=0.02\,\mathrm{eV}$. Other parameters: $r_D=10^8\,\mathrm{km}$, $r_S=10^5r_D$, $E_0=10\,\mathrm{MeV}$ and $|\Delta m^2|=10^{-3}\,\mathrm{eV^2}$.}
\end{figure*}

We proceed the numerical discussion on the neutrino two-flavour toy model. The neutrino flavour conversion spanning the azimuthal angle $\phi\in[0, 0.003]$ is shown in Figs.\ref{fig:prob1}, \ref{fig:prob2} and \ref{fig:prob3}. In Fig.\ref{fig:prob1}, the neutrino oscillation probability $\nu_e\to \nu_{\mu}$ is displayed for $\Lambda=0.02$ (first and second panels) and $\Lambda=0.04$ (third and fourth panels). The positive level $\left(\Delta m^2>0\right)$ is represented with the lower line, while the inverse level $\left(\Delta m^2<0\right)$ is represented by with upper line. The mixing angles are $\alpha=\frac{\pi}{5}$ and $\alpha=\frac{\pi}{6}$ and the inverted ordering is always correlated with a higher conversion probability than the normal ones. Upon initial examination of these four figures as parts of Fig.\ref{fig:prob1}, it is evident that the oscillation probabilities of normal and inverted mass orderings differ significantly, with the exception of a small number of $\phi$ values. This implies that the nature of $\Delta m^2$ influences on the wormhole lensing drastically. On the other hand, the two-flavour vacuum oscillation probabilities in flat spacetime have no sign difference in $\Delta m^2$ \cite{swami2020signature}. The same findings are displayed in Fig.\ref{fig:prob2} to further illustration on how the probabilities depend on $\Lambda$. When the factor $\Lambda$ becomes smaller, the whole conversion probabilities will be larger and the probabilities approach with larger values of $\Lambda$, which is similar to Ref.\cite{chakrabarty2022effects}. According to the four parts of the Fig.\ref{fig:prob2}, the curves for $\Delta m^2$ with the same sign have similar forms for some values of $\Lambda$. The Damour-Solodukhin parameter $\Lambda$ affects the oscillation probabilities of the neutrinos $\nu_e$ and $\nu_{\mu}$ clearly. To examine this, we have rearranged the curves in Fig.\ref{fig:prob2} that correspond to $\Lambda$, the mixing angle $\alpha$, and the symbols of $\Delta m^2$. The magnitudes of probability functions associated with the azimuthal angles differ for factor $\Lambda$ with different values although the curve forms with the same mixing angle $\alpha$ and the same sign of $\Delta m^2$ are similar. The values of $\mathcal{P}_{\alpha\beta}^{\mathrm{lens}}$ for various magnitudes of $\Lambda$ are different in Fig. \ref{fig:prob2} when there is a triple covariance (e.g., $\phi= 0$) while the neutrino source, the wormhole that resembles a black hole, and the detector are colinear. The detailed comparisons among the numerical estimation of $\nu_{e}\to\nu_{\mu}$ conversion are made and it is evident that the probability curves for Damour-Solodukhin factor with different values do not coincide each other.

For the cases of mixing angles $\alpha=\frac{\pi}{5}$ and $\alpha=\frac{\pi}{6}$ respectively, the first two images in Fig.\ref{fig:prob2} show the transition probabilities for $\Lambda=0$(upper line), $\Lambda=0.02$(middle line), and $\Lambda=0.04$(lower line) for the normal hierarchy $\left(\Delta m^2>0\right)$. Here the lightest neutrinos are considered to be massless in Figs.\ref{fig:prob1} and \ref{fig:prob2}. The final two images depict the transition probabilities for the inverse hierarchy $\left(\Delta m^2<0\right)$, with the three curves signifying the transition probabilities for $\Lambda=0$, $\Lambda=0.02$, and $\Lambda=0.04$ respectively. The transition probabilities at the normal and inverted orderings both decrease as $\Lambda$ grows within the range $\phi$.

The transition probability for the lightest neutrino including the massless and massive cases is shown in Fig.\ref{fig:prob3}. The conversion probabilities for the normal ordering with $\Lambda=0.02$ and $0.04$ for the lightest neutrino masses, $m_1=0\,\mathrm{eV}$, $0.01\,\mathrm{eV}$, and $0.02\,\mathrm{eV}$ when $\Delta m^2>0$, are demonstrated in the first two panels of Fig.\ref{fig:prob3} and the third and fourth plots of Fig.\ref{fig:prob3} depict the inverted ordering scenario. It is clear that the curves subject to the individual neutrino masses are not coincident under the same-valued parameter $\Lambda$. In each panel with $\alpha=\frac{\pi}{5}$ or $\alpha=\frac{\pi}{6}$ and $\Delta m^{2}>0$ or $\Delta m^{2}<0$, the probability for the neutrino flavour oscillation with respect to the azimuthal angle $\phi$ depends on the individual mass of neutrino, which is similar to the case of Schwarzschild black holes \cite{swami2020signature}. The probability differences due to the difference magnitude of $\Lambda$ belonging to the Damour-Solodukhin wormhole can be significantly measurable.

\subsection{Three flavor case}

We investigate the neutrino lensing effect due to the Damour-Solodukhin wormholes in the case of three flavours are presented in this section. We study the transitions $\nu_e\to\nu_{\mu}$, $\nu_e\to\nu_{\tau}$, and $\nu_{\mu}\to\nu_{\tau}$ from the oscillatory probability expressions corresponding to to Eq.(\ref{Pab_lens}). A popular PMNS matrix is parametrized with the Dirac CP phase $\delta_{CP}$ and the three angles $\theta_{12}$, $\theta_{13}$, and $\theta_{23}$ \cite{capozzi2014status,de2018status,esteban2019global}. We utilize neutrino oscillation data (with SK atmospheric data including the neutrino mass and mixing parameters \cite{esteban2020fate,an2012observation}. The parameters are $\Delta m_{21}^2=7.36\times10^{-5}\,\mathrm{eV^2}$, $\Delta m_{31}^2=2.519\times10^{-5}\,\mathrm{eV^2}\left(\Delta m_{21}^2=-2.497\times10^{-5}\,\mathrm{eV^2}\right)$, $\theta_{12}=33.32^{\circ}\left(\theta_{12}=33.40^{\circ}\right)$, $\theta_{13}=8.69^{\circ}\left(\theta_{13}=8.72^{\circ}\right)$, $\theta_{23}=49.9^{\circ}\left(\theta_{23}=49.8^{\circ}\right)$ and $\delta_{CP}=217^{\circ}\left(\delta_{CP}=297^{\circ}\right)$ for normal (inverted) ordering \cite{esteban2020fate,an2012observation}.

\begin{figure*}
\centering
\includegraphics[height=5cm]{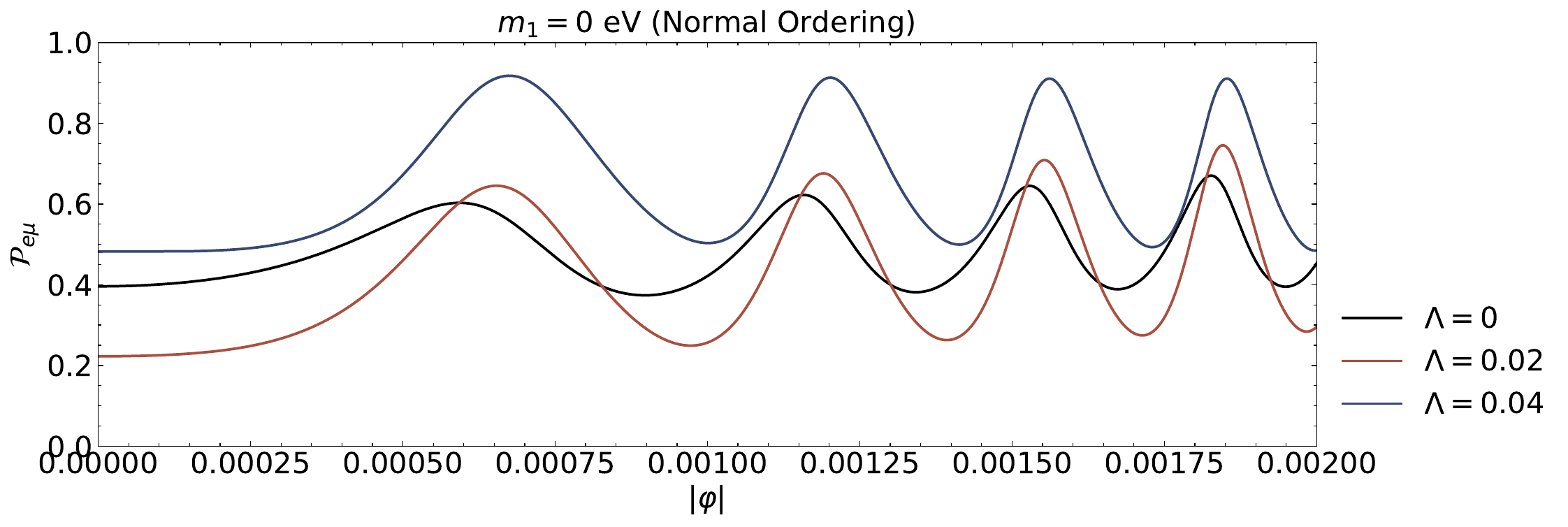}
\includegraphics[height=5cm]{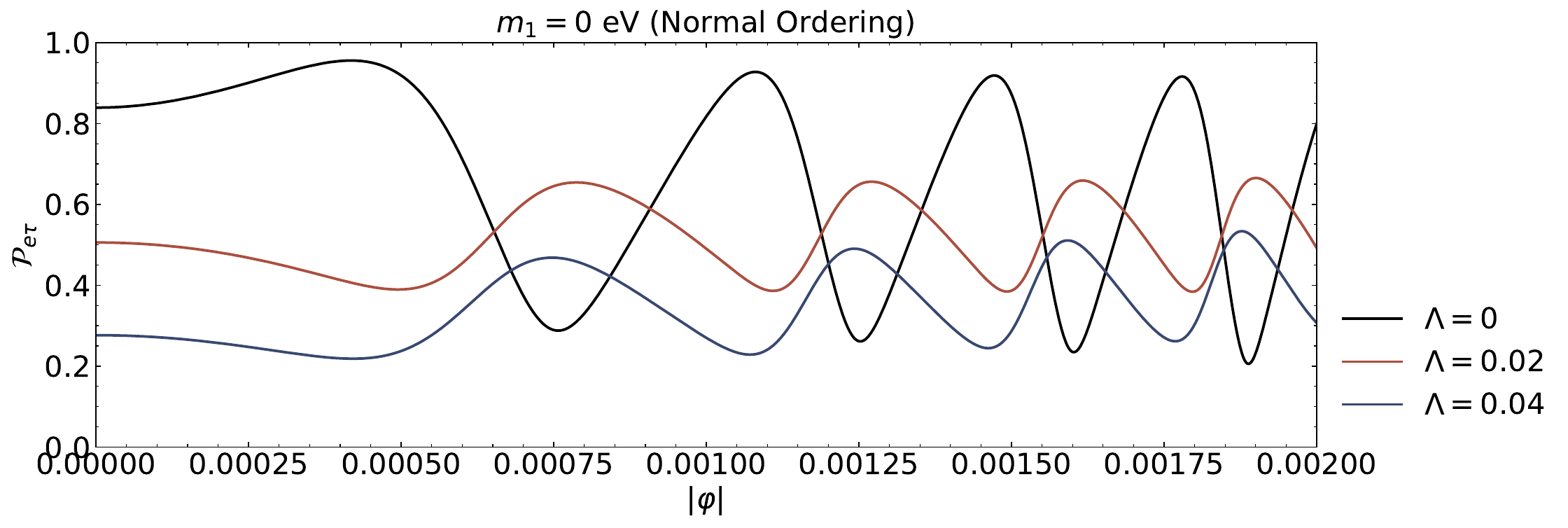}
\includegraphics[height=5cm]{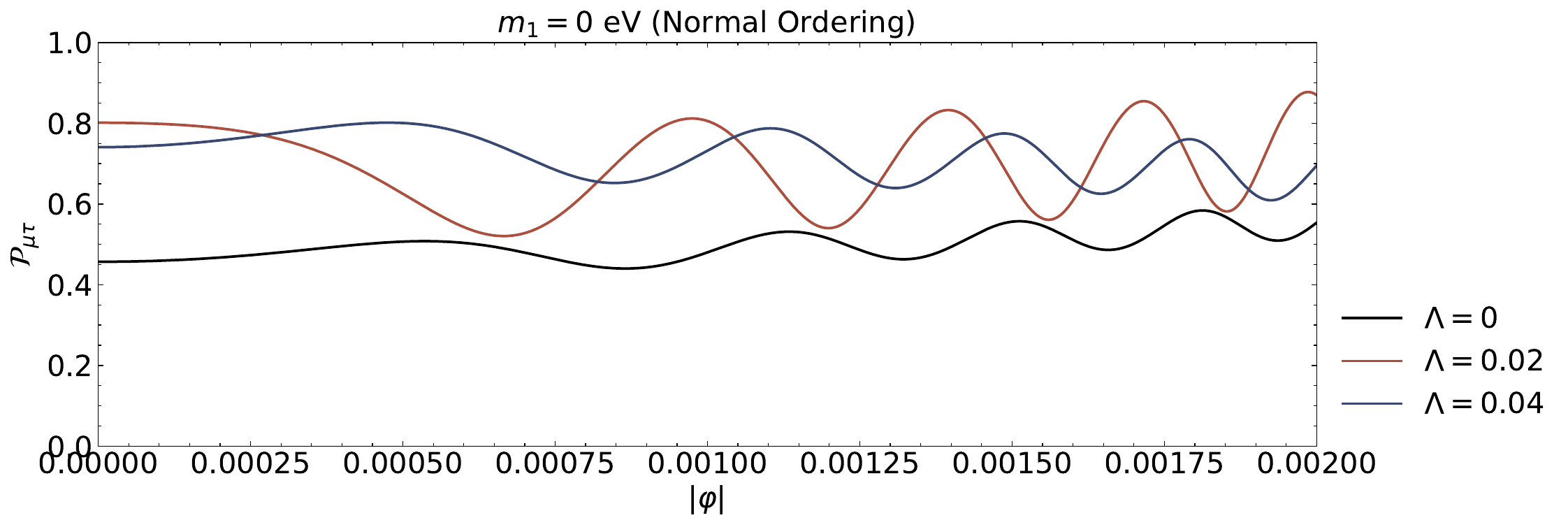}
\caption{\label{fig:prob4} Probabilities of oscillation in the normal hierarchy for the three-flavor situation when $m_1=0\,\mathrm{eV}$. Upper panel: the probability of the conversion $\nu_e\to\nu_{\mu}$ when $\Lambda=0$ (black line), $\Lambda=0.02$ (red line), and $\Lambda=0.04$ (blue line). Middle panel: For $\Lambda=0$ (black line), $\Lambda=0.02$ (red line), and $\Lambda=0.04$ (blue line), the probability of the conversion $\nu_e\to\nu_{\tau}$ is shown. Lower panel: For $\Lambda=0$ (black line), $\Lambda=0.02$ (red line), and $\Lambda=0.04$ (blue line), the probability of the conversion $\nu_{\mu}\to\nu_{\tau}$. We assume that $r_D=10^8\,\mathrm{km}$, $r_S=10^5r_D$, and $E_0=10\,\mathrm{MeV}$.}
\end{figure*}

\begin{figure*}
\centering
\includegraphics[height=5cm]{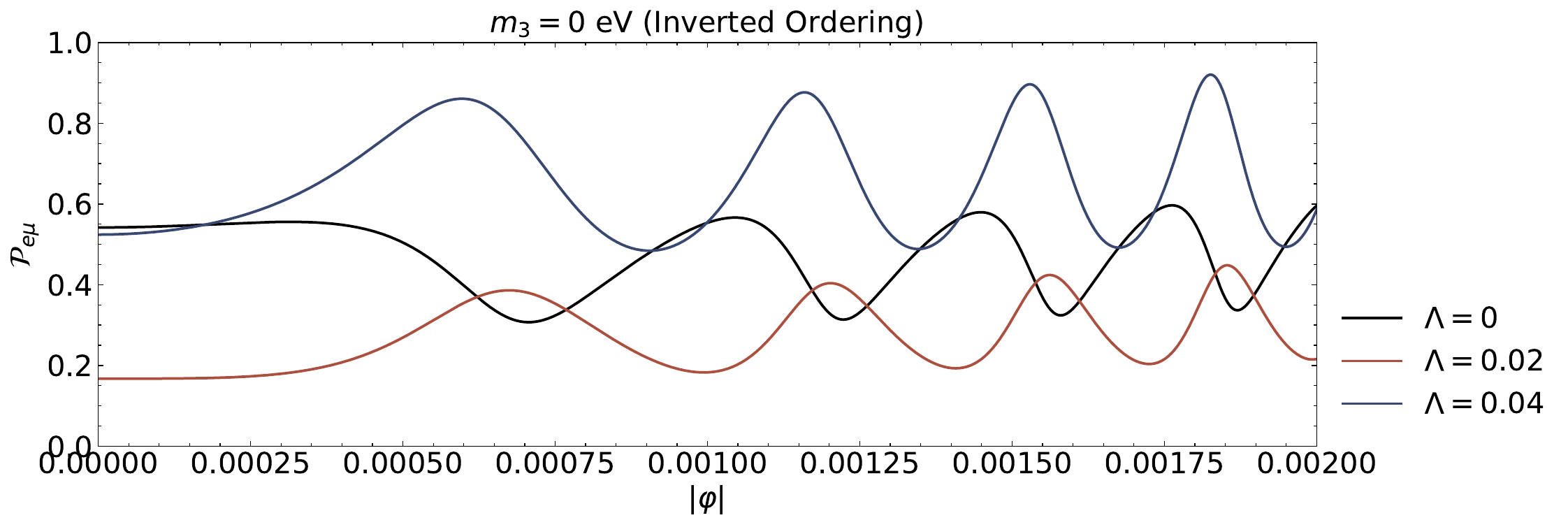}
\includegraphics[height=5cm]{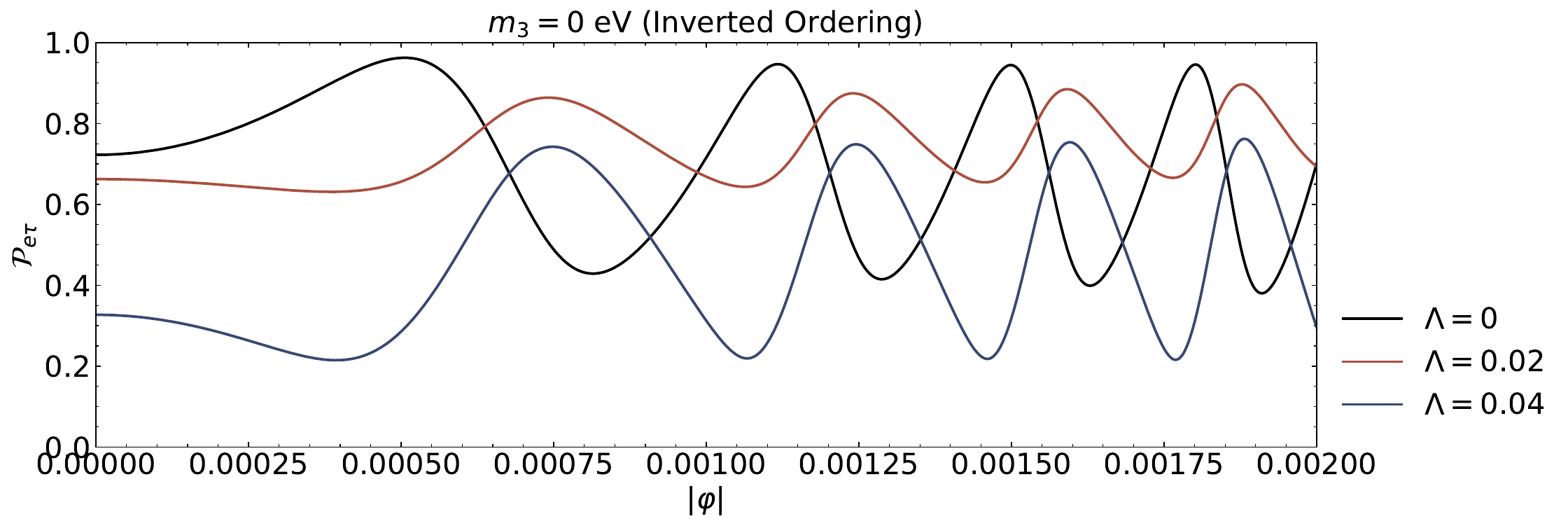}
\includegraphics[height=5cm]{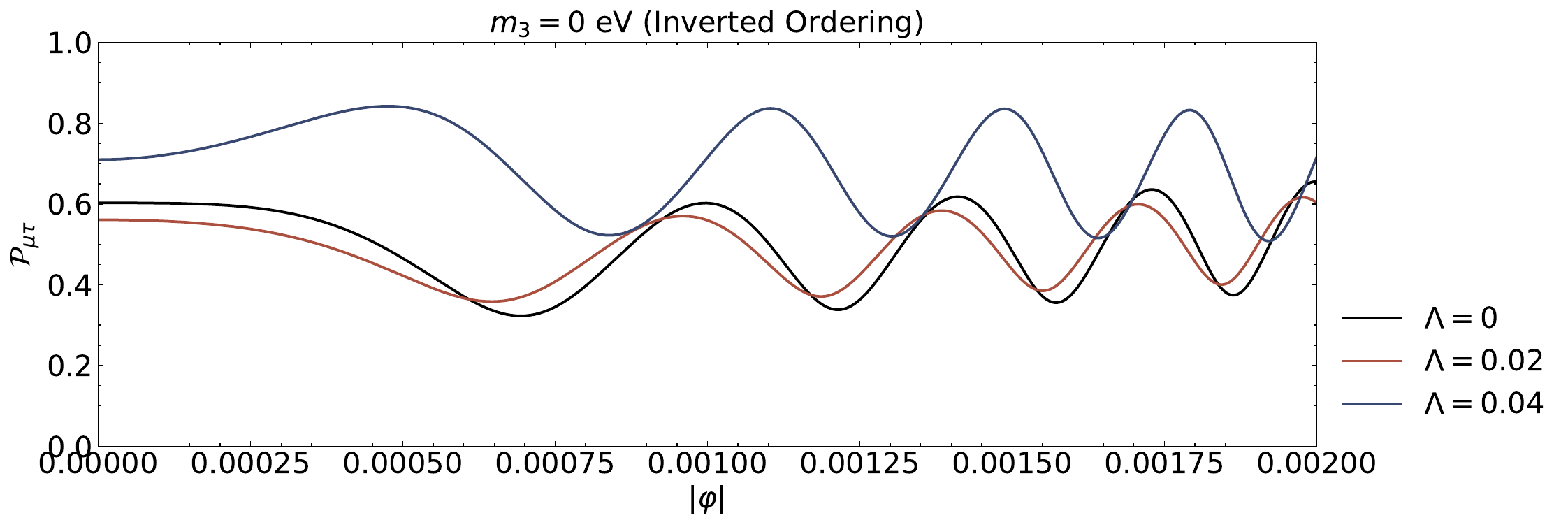}
\caption{\label{fig:prob5} Probabilities of oscillation in the inverted hierarchy for the three-flavor situation when $m_3=0\,\mathrm{eV}$. Upper panel: the probability of the conversion $\nu_e\to\nu_{\mu}$ when $\Lambda=0$ (black line), $\Lambda=0.02$ (red line), and $\Lambda=0.04$ (blue line). Middle panel: For $\Lambda=0$ (black line), $\Lambda=0.02$ (red line), and $\Lambda=0.04$ (blue line), the probability of the conversion $\nu_e\to\nu_{\tau}$ is shown. Lower panel: For $\Lambda=0$ (black line), $\Lambda=0.02$ (red line), and $\Lambda=0.04$ (blue line), the probability of the conversion $\nu_{\mu}\to\nu_{\tau}$. We assume that $r_D=10^8\,\mathrm{km}$, $r_S=10^5r_D$, and $E_0=10\,\mathrm{MeV}$.}
\end{figure*}

The transition probability as a function of the azimuthal angle $\phi = 0$ in the presence of Damour-Solodukhin factor $\Lambda$ is depicted in the Figs.\ref{fig:prob4} and \ref{fig:prob5}. The wormhole factor can change the curve shapes of the function more greatly. We can compare our results with the observations to judge whether the remote gravitational source is black hole or black hole-like wormhole. In the case of three flavours of neutrinos, the sign of squared mass difference and the individual mass with its own mixing angle also modify the lensing probability function of neutrino oscillation on the azimuthal angle.

\section{Conclusion}

We investigate the geodesic in the background governed by the Damour-Solodukhin metric to describe the propagation of neutrinos in the cases of two and three flavours of neutrinos respectively. The discussions are performed to the gravitational lensing of neutrinos that the neutrinos setting free from the emitter go through the black-hole-like wormhole and converge to the detector. We also derive the phase connecting the mass eigenstate and the flavour eigenstate and the phase difference for all paths that the neutrinos propagate along in the curved spacetime. The conversion probabilities for the neutrino lensing owing to the Damour-Solodukhin wormhole have been computed in the cases of two-flavoured neutrinos or three-flavoured ones. The relationships between the oscillation probability associated with neutrino masses and the structure of black-hole-like wormhole are displayed graphically. We find that the oscillation probabilities for the wormholes involving the deviation from the Schwarzschild black holes also have something to do with the sign of difference of squared neutrino masses no matter how strong the deviation is, which is similar to the outcome in Ref.\cite{cardall1997neutrino}. The spacetime around the gravitational source with a throat is not flat but curved and we discover that the flavour conversion also relate to the individual masses of neutrinos. We have paid our efforts for the researches on the Damour-Solodukhin-wormhole-induced effects on the neutrino oscillation to declare the properties and the wormhole feature conforms to those of black hole \cite{cardall1997neutrino}.

It should be pointed out that the deviation from the regular black hole brings about the considerable effects on the many aspects of the neutrino oscillation. Based on our researches, we find that the influence from Damour-Solodukhin factor on the neutrino oscillation is measurable. If the factor $\Lambda$ shifts a little, the probability magnitude and the corresponding curve shapes will change evidently for the definite quantities such as mixing angle, absolute masses of neutrinos. We can adjust the parameter $\Lambda$ to compute and plot a series of curves of neutrino conversion probability versus the azimuthal angle theoretically. It is possible to compare our numerical results with the measurements of flavour transition probabilities for a distant celestial body to to determine whether the gravitational source is a black hole with $\Lambda=0$ or a wormhole with a shift in a black hole with $\Lambda>0$. The fine distinctions among the corrections from the wormhole parameter on the neutrino flavour oscillation can be distinguished in flavour of our calculations and analysis. Further we can estimate the parameter $\Lambda$ in this research. We have found that there is a direct relationship between the wormhole structure and the neutrino oscillation. This work can be thought as a new window to explore and confirm the static and spherically black-hole-like wormhole. We can generalize the technique to the other gravitational sources.

\vspace{1cm}
\noindent \textbf{Acknowledge}

This work is partly supported by the Shanghai Key Laboratory of
Astrophysics 18DZ2271600.

\newpage
\bibliography{reference}

\end{document}